\documentclass[onecolumn,epjc3]{svjour3} 
 
\usepackage[T1]{fontenc}
\usepackage{epsfig,amsmath,amssymb}
\usepackage{graphicx}
\usepackage{xspace}
\usepackage{listings}
\usepackage{ulem}
\usepackage{slashed}
\usepackage{cite}
\usepackage{nicefrac}
\usepackage{feynmp}
\usepackage[colorlinks,citecolor=blue,urlcolor=blue,linkcolor=blue]{hyperref}
\journalname{Eur. Phys. J. C}

\def\hc{\text{h.c.}}

\newcommand\SARAH{{\tt SARAH}\xspace}

\newcommand\SPheno{{\tt SPheno}\xspace}

\newcommand\FlavorKit{{\tt FlavorKit}\xspace}

\newcommand\Mathematica{{\tt Mathematica}\xspace}

\newcommand{\Fortran}{\texttt{Fortran}\xspace}

\newcommand{\newc}{\newcommand}
\newcommand{\qref}[1]{{eq.~(\ref{#1})}}
\newc{\Psibar}{\overline{\Psi}}
\newc{\FFbS}{\overline{FF}S}
\newc{\FFbV}{\overline{FF}V}
\newc{\FFbe}{\overline{FF}\epsilon}
\newc{\FSS}{F_{SS}}
\newc{\FSSS}{F_{SSS}}
\newc{\FFFS}{F_{FFS}}
\newc{\FFFbS}{F_{\overline{FF}S}}
\newc{\FSSV}{F_{SSV}}
\newc{\FVS}{F_{VS}}
\newc{\FVVS}{F_{VVS}}
\newc{\FFFV}{F_{FFV}}
\newc{\FFFbV}{F_{\overline{FF}V}}
\newc{\FVV}{F_{VV}}
\newc{\FVVV}{F_{VVV}}
\newc{\fggV}{f_{ggV}}
\newc{\Fgauge}{F_{\rm gauge}}
\newc{\fSS}{f_{SS}}
\newc{\fSSS}{f_{SSS}}
\newc{\fFFS}{f_{FFS}}
\newc{\fFFbS}{f_{\overline{FF}S}}
\newc{\fSSV}{f_{SSV}}
\newc{\fVVS}{f_{VVS}}
\newc{\fVS}{f_{VS}}
\newc{\fFFV}{f_{FFV}}
\newc{\fFFbV}{f_{\overline{FF}V}}
\newc{\fVV}{f_{VV}}
\newc{\fVVV}{f_{VVV}}
\newc{\fgauge}{f_{\rm gauge}}
\newc{\FFVbar}{{\overline{FF}V}}
\newc{\FFSbar}{\overline{FF}S}

\newcommand{\spa}[1]{\widetilde #1}

\newcommand{\abs}[1]{\left|#1\right|}

\newcommand{\kl}[1]{\left(#1\right)}
\newcommand{\nn}{\nonumber}

\newcommand{\delup}[1]{\partial^{#1}}
\newcommand{\str}{\prime}
\newc{\bsigmu}{\bar\sigma^\mu}
\newc{\DRbar}{\overline{\text{DR}}}
\newc{\DRbarprime}{\overline{\text{DR}}^\prime}
\newcommand{\delmulr}{\overleftrightarrow{\partial^\mu}} 

\begin{document}

\title{Two-Loop Higgs mass calculations in supersymmetric models beyond the MSSM with \SARAH and \SPheno}
\author{
   Mark D. Goodsell \thanksref{a1} \and
   Kilian Nickel \thanksref{a2} \and
   Florian Staub \thanksref{a2,a3}
   }

\institute{ Sorbonne Universit\'es, UPMC Univ Paris 06, UMR 7589, LPTHE, F-75005, Paris, \& \\
CNRS, UMR 7589, LPTHE, F-75005, Paris, France\label{a1}
\and
BCTP \& Physikalisches Institut der 
Universit\"at Bonn, Nu{\ss}allee 12, 53115 Bonn, Germany\label{a2}
\and
Theory Division, CERN, 1211 Geneva 23, Switzerland\label{a3}
}

\date{BONN-TH-2014-14, CERN-PH-TH-2014-213}

\maketitle

\lstset{frame=shadowbox}
\lstset{prebreak=\raisebox{0ex}[0ex][0ex]
        {\ensuremath{\hookrightarrow}}}
\lstset{postbreak=\raisebox{0ex}[0ex][0ex]
        {\ensuremath{\hookleftarrow\space}}}
\lstset{breaklines=true, breakatwhitespace=true}
\lstset{numbers=left, numberstyle=\scriptsize}
 
\begin{abstract}
We present an extension to the Mathematica package \SARAH which 
allows for Higgs mass calculations at the two-loop level in a wide range 
of supersymmetric (SUSY) models beyond the MSSM. These calculations are based on 
the effective potential approach and 
include all two-loop corrections which 
are independent of electroweak gauge couplings.
For the numerical evaluation 
Fortran code for \SPheno is generated by \SARAH. 
This allows the prediction of
the Higgs mass in more complicated SUSY models with the same precision that most state-of-the-art 
spectrum generators provide for the MSSM.
\end{abstract}

% \tableofcontents

\section{Introduction}
The discovery of the Higgs boson has been so far the biggest success of the experiments at the Large Hadron Collider (LHC) \cite{Chatrchyan:2012ufa,Aad:2012tfa}. The mass of the Higgs is already pinned down with an impressive experimental uncertainty of just a few hundred MeV in the range of 125 -- 126~GeV. This experimental accuracy is at the moment much better than theoretical predictions for the Higgs mass in any given model beyond the standard model (SM). For instance, in recent decades a lot of effort has been taken to calculate the Higgs mass in the minimal supersymmetric standard model (MSSM). This industry was initiated by the observation that stop corrections can lift the Higgs mass, which is bounded at tree-level to be below $M_Z$, above the long existing LEP limit of 114~GeV \cite{Haber:1990aw,Ellis:1990nz,Okada:1990vk,Okada:1990gg,Ellis:1991zd}. The next milestones were a complete diagrammatic one-loop calculation \cite{Brignole:1992uf,Chankowski:1991md,Dabelstein:1994hb,Pierce:1996zz} and a calculation of the leading two-loop corrections in 
the effective potential approach \cite{Hempfling:1993qq,Carena:1995wu,Heinemeyer:1998jw,Zhang:1998bm,Heinemeyer:1998np,Heinemeyer:1999be,Espinosa:1999zm,Espinosa:2000df,Brignole:2001jy,Degrassi:2001yf,Martin:2002wn,Brignole:2002bz,Dedes:2002dy,Martin:2002iu,Dedes:2003km} or equivalent diagrammatic calculations 
with zero external momentum \cite{Heinemeyer:2004xw} on the one side as well as progressively better calculations using renormalisation group equation (RGE) methods \cite{Sasaki:1991qu,Carena:1995bx,Haber:1996fp,Carena:2000dp,Carena:2000yi,Carena:2001fw,Espinosa:2001mm}. In particular the two-loop calculation in the effective potential including $O\left(\alpha_s (\alpha_t+\alpha_b+\alpha_\tau)\right)$ and $O\left((\alpha_t + \alpha_b + \alpha_\tau)^2\right)$ corrections are widely used because they have entered different public codes such as {\tt SoftSUSY} \cite{Allanach:2001kg,Allanach:2009bv,Allanach:2014nba}, {\tt SPheno} \cite{Porod:2003um,Porod:2011nf}, {\tt Suspect} \cite{Djouadi:2002ze} or {\tt FeynHiggs} \cite{Heinemeyer:1998yj,Hahn:2009zz}. The RGE methods are implemented in {\tt CPsuperH} \cite{Lee:2003nta,Ellis:2006eh}. Also three-loop results in the effective potential approach exist 
\cite{Martin:2007pg,Kant:2010tf,Harlander:2008ju}. The discovery of the Higgs and the determination of its mass gave a new impetus to these calculations and now the diagrammatic two-loop calculations $O(\alpha_s \alpha_t)$ and $O(\alpha_t^2)$  including external momenta exist \cite{Martin:2004kr,Borowka:2014wla,Hollik:2014bua,Degrassi:2014pfa}. In addition, calculations in an effective model  matched to the MSSM at a higher scale have been performed \cite{Draper:2013oza,Bagnaschi:2014rsa}.  However, if one goes beyond the MSSM and considers non-minimal SUSY models, the picture is very simple: only for the next-to-minimal supersymmetric standard model (NMSSM) a full one-loop calculation has been presented in the literature \cite{Degrassi:2009yq,Staub:2010ty,Ender:2011qh,Graf:2012hh}. At the two-loop level only the $O(\alpha_s (\alpha_t + \alpha_b))$ in the effective potential approach are known up to now \cite{Degrassi:2009yq}. This is in particular problematic because non-minimal SUSY models 
have gained more and more interest in the last few years because of the experimental results: (i) they can lift the Higgs mass at tree-level by new $F$- or  $D$-contributions \cite{Ellwanger:2009dp,Ellwanger:2006rm,Ma:2011ea,Zhang:2008jm,Hirsch:2011hg} which makes these models more natural by reducing the fine-tuning \cite{BasteroGil:2000bw,Dermisek:2005gg,Ellwanger:2011mu,Ross:2011xv,Ross:2012nr}; (ii) they can weaken SUSY limits by either predicting compressed spectra or reducing the expected missing transverse energy significantly \cite{Dreiner:2012gx,Bhattacherjee:2013gr,Ellwanger:2014hia,Kim:2014eva}. A brief overview of Higgs sectors in BMSSM models is given, for instance, 
in \cite{Staub:2014pca}. 

The situation in non-minimal SUSY models has been relaxed with the development of the Mathematica package \SARAH \cite{Staub:2008uz,Staub:2009bi,Staub:2010jh,Staub:2012pb,Staub:2013tta}: \SARAH can automatically generate \SPheno modules which allow for a full one-loop calculation in a wide range of SUSY models like singlet extensions \cite{Ross:2012nr,Louis:2014pia,Kaminska:2014wia}, triplet extensions \cite{Arina:2014xya}, models with Dirac gauginos and broken \cite{Benakli:2012cy,Benakli:2014cia}
or unbroken \cite{Diessner:2014ksa} $R$-symmetry, extended gauge sectors \cite{O'Leary:2011yq,Hirsch:2011hg,Hirsch:2012kv,Brooijmans:2014eja} or even more exotic models \cite{Banerjee:2013fga,Bharucha:2013ela}. A similar link between \SARAH and {\tt FlexibleSUSY} has been presented recently \cite{Athron:2014yba}. 
These automatised calculations have now been brought to the next level by providing routines which calculate two-loop  corrections to the CP even Higgs scalars in the effective potential approach. The resulting accuracy for many beyond-MSSM models is the same as for the MSSM using the results of Refs.~\cite{Brignole:2001jy,Degrassi:2001yf,Brignole:2002bz,Dedes:2002dy,Dedes:2003km}. For this purpose, the generic results for the two-loop effective potential presented in Ref.~\cite{Martin:2001vx} have been implemented in \SARAH and the two-loop self-energies in the approximation of vanishing external momenta are calculated using a numerical derivation. This is analog to Ref.~\cite{Martin:2002wn}. Corrections including gauge couplings of a broken gauge group are not included.

This paper is organized as follows. In sec.~\ref{sec:method} we review the effective potential approach as well as our implementation in \SARAH and \SPheno. In sec.~\ref{sec:validation} we compare the results obtained for the MSSM and NMSSM with the results of well established routines. In sec.~\ref{sec:usage} we explain how the user can obtain the two-loop results for his/her favorite model before we conclude in sec.~\ref{sec:conclusion}.

\section{Effective potential approach for two-loop Higgs masses in a generic SUSY model}
\label{sec:method}
We consider a set of real scalars $\{\phi_k\}$ which are diagonalised to  
physical states $\{h_k\}$. The scalar potential at tree-level is $V^T(\{\phi_k\})$. The pole masses of $h_i$ are in general the eigenvalues of the 
loop corrected mass matrix given by
\begin{equation}
\frac{\partial^2 V^T}{\partial \phi_i \partial \phi_j} + \sum_n \Pi^{(n)}_{ij}(p^2)
\end{equation}
Here, $\Pi^{(n)}(p^2)$ is the self-energy at the $n$-loop level which usually depends on the 
external momenta. The parameters appearing in $V^T$ should be chosen to minimise the 
effective potential. For a pole mass $m_{h_i}$ the momentum is fixed to be $p^2 = m_{h_i}^2$. To include 
this momentum 
dependence of $\Pi^{(n)}$ a diagrammatic calculation is necessary. However, things become significantly easier if one 
considers the limit of $p^2=0$. In this limit $\Pi^{(n)}(0)$ is equivalent to the second derivative of the 
effective potential at the $n$-loop level,
\begin{equation}
\label{eq:effpotApprox}
\Pi_{ij}^{(n)}(0)  = \frac{\partial^2 V^{(n)}}{\partial \phi_i \partial \phi_j} \, .
\end{equation}
We present here a fully automatised hybrid method implemented in the public tools \SARAH and \SPheno 
for the calculation of the scalar masses at the two-loop 
level in the $\DRbarprime$ renormalization scheme \cite{Jack:1994rk}: while the one-loop corrections are calculated including 
the full momentum dependence, the two-loop corrections are derived in the effective potential approximation. 
In general, the setup is based on the following work distribution: the user implements their favourite model in \SARAH. 
\SARAH derives all analytical expressions for mass-matrices, vertices, renormalisation group equations as well as 
loop corrections and exports 
this information into \Fortran source code. The \Fortran source code is compiled together with \SPheno and all numerical 
calculations are then performed by the new \SPheno module. 
Since the one-loop diagrammatic calculation has been included since \SARAH\ {\tt 2.0}, we focus in the following on the new two-loop 
corrections which are published in version {\tt 4.4.0}. We start with a discussion how the two-loop self-energies are derived.

\subsection{Calculation of the two-loop self energies in the effective potential approximation}
\begin{figure}[hbt]
\begin{center}
\includegraphics[width=0.8\linewidth]{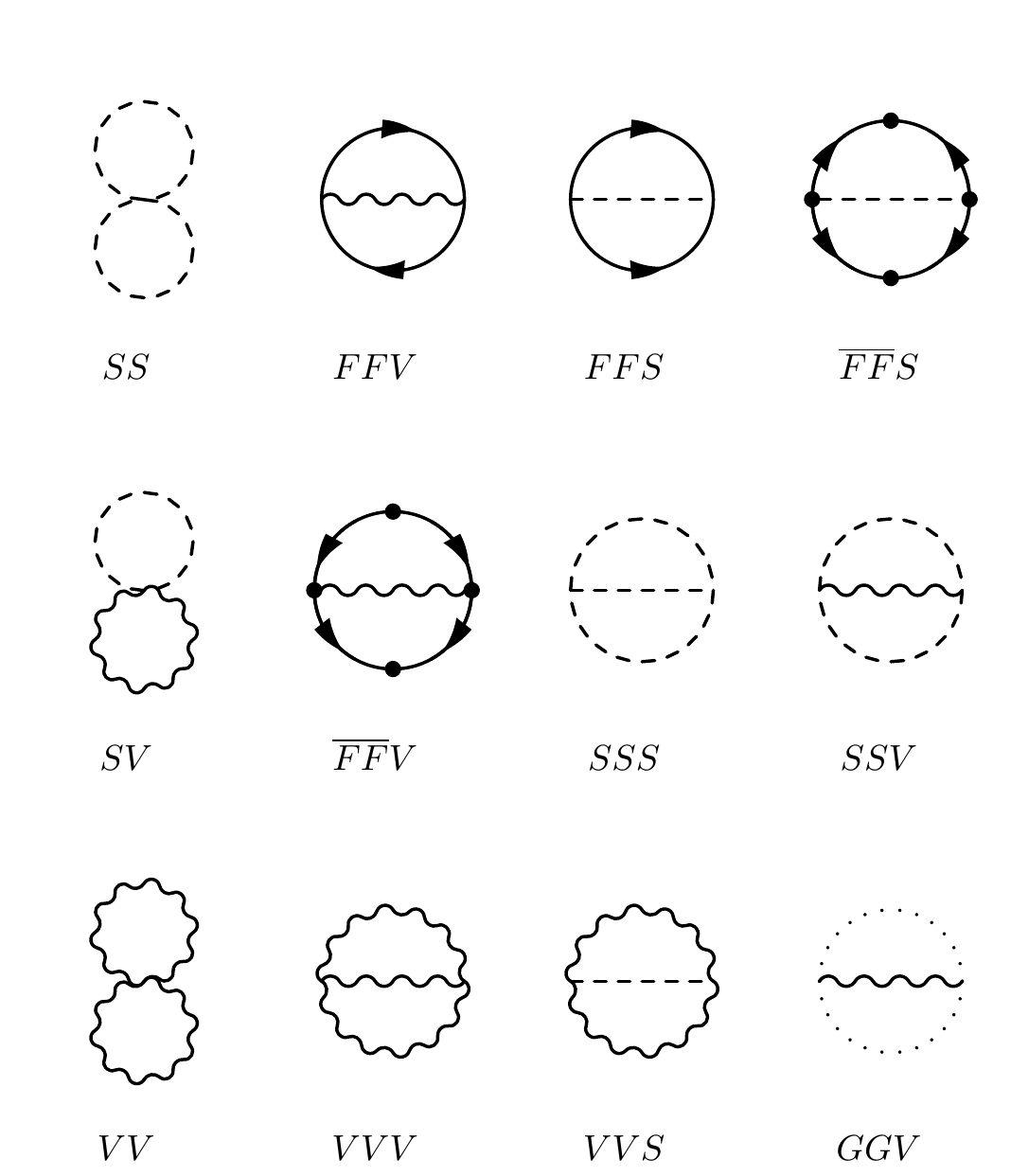}
\caption{All possible topologies of two-loop bubble diagrams. We consider in the following 
$SS$, $SSV$, $SSS$, $FFS$, $\overline{FF}S$, $FFV$, and $\overline{FF}V$. 
}
\label{fig:topologies}
\end{center}
\end{figure}

We shall neglect CP-violating effects in the following. Therefore, a set of neutral, complex scalars $H_i$ are decomposed after 
symmetry breaking as
\begin{equation}
H_i = \frac{1}{\sqrt{2}}\left(\phi_i + i \sigma_i + v_i \right) \, .
\end{equation}
$\phi_i$ are the CP-even components, $\sigma_i$ the CP-odd ones and $v_i$ are the vacuum expectation values (VEVs). 
Under this assumption, the two-loop corrections to the mass matrix of real scalars in the effective potential approach are given by
\begin{equation}
\label{eq:Pi2L}
\Pi^{(2)}_{ij}(0) = \frac{\partial^2 V^{(2)}}{\partial v_i \partial v_j}\,.
\end{equation}
In addition, we require the  tadpole contributions which are the first derivatives 
of the effective potential,
\begin{equation}
\label{eq:Ti2L}
\delta t_i^{(2)} = \frac{\partial V^{(2)}}{\partial v_i} \,.
\end{equation}
The tadpoles then also contribute to the scalar masses by shifting the parameters; for example if we treat the vevs $v_i$ as fixed, ``all-loop'' correct values, then we can exchange them for scalar mass-squared at a given loop order via the tadpole equations, and in turn find compact equations for the total shift in the mass-squareds of
\begin{align}
\big( \Delta \mathcal{M}^2_S \big)_{ij}^{\rm eff} =& \frac{\partial^2 V^{(2)}}{\partial v_i \partial v_j} -\frac{\delta_{ij}}{v_i}  \frac{\partial V^{(2)}}{\partial v_i },
\end{align}
as used, for example, in \cite{Degrassi:2009yq}. However, \emph{we do not do this here}, instead (as detailed below) solving the tadpole equations and then using the parameters derived from these in the tree-level mass calculation. The reason is that \SARAH allows for a more general choice of variables to solve for via the tadpole equations.

$V^{(2)}$ receives contributions from the possible topologies shown in Fig.~\ref{fig:topologies}. The generic results for all of these diagrams in Landau gauge have been presented in Ref.~\cite{Martin:2001vx} and we heavily make use of these 
results in the following. However, we will not require all topologies but neglect $VV$, $VVS$, $VVV$, $SV$, and $GGV$. The reason is that 
these only lead to non-zero contributions if massive vector bosons are involved. 
We will neglect all contributions  stemming from
broken gauge groups because of the many complications that they entail. 

The remaining topologies are those which have been considered so far in the MSSM to obtain the dominant 
two-loop corrections and which are implemented in public computer tools. The impact 
on the Higgs mass is considered to be moderate when neglecting 
contributions propotional to the electroweak gauge couplings in models with the SM gauge sector, since they are significantly smaller than the strong and top Yukawa coupling -- and the diagrams involving these couplings are of typically lower multiplicity. In fact, by inspecting the form of the loop functions it is evident that the topologies $SV, VV, VVV,VVS $ and $GGV$ lead to contributions proportional to $g^4_{EW} v^2 $ where $g_{EW}$ stands for the electroweak gauge coupling; these are hence subdominant to the contributions from $SSV, FFV, \overline{FF}V$ diagrams involving the electroweak gauge bosons (which we are also neglecting) which would be proportional to $g_{EW}^2 v^2 $. 

The electroweak contributions 
in the MSSM have been estimated to be of $O(1~\text{GeV})$ \cite{Degrassi:2002fi}. A more recent estimate of 
the total theoretical uncertainty of the MSSM Higgs mass concluded that missing two- and three-loop contributions 
together can account for a shift of about 2~GeV in the Higgs mass if third generation squarks
are 2 TeV or lighter \cite{Buchmueller:2013psa}. This is much smaller 
than the corrections involving superpotential interactions and the strong coupling. 
Of course, in models with extended gauge sectors this conclusion might change. An estimate of the 
importance of the missing contributions can be obtained be considering the different one-loop corrections
and assuming a similar behavior at two loops.

Note, for consistency, we  also set all gauge couplings arising in the $D$-terms of broken groups in the vertices to zero. 
The masses used in the loops are  tree-level masses calculated from running $\DRbarprime$ parameters. 
In this context, there are two possibilities to treat $D$-term contributions to the tree-level masses: either one can work in the gaugeless limit 
 where these contributions are also put to zero \cite{Degrassi:2001yf}, or one can work with the full tree-level masses as this is also 
 done in diagrammatic calculations \cite{Borowka:2014wla}. We  offer 
both possibilities.
In the gaugeless limit the $SU(2)_L \times U(1)_Y$ gauge symmetries become global symmetries. Therefore, it is obvious that 
no gauge dependence 
has been introduced by including only Goldstone diagrams but no diagrams of the corresponding vector bosons.
 Of course, when including $D$-terms in the mass matrices 
the derivatives of the $D$-terms are still forced to vanish. However, 
we stress that the second method has to be used carefully as explained in sec.~\ref{sec:goldstones}.

For the calculation of the two-loop effective potential we have translated the expressions of Ref.~\cite{Martin:2001vx} given in two-component notation
into four-component language, see  appendix~ \ref{app:FourComponentNotation}. All necessary generic expressions have been implemented in  \SARAH. 
\SARAH uses these expressions to generate \Fortran code for all two-loop diagrams which are possible in the considered model
assuming the topologies $SS$, $SSV$, $SSS$, $FFS$, $\overline{FF}S$, $FFV$ and $\overline{FF}V$.

As soon as the two-loop effective potential is calculated, one can obtain the two-loop corrections to the Higgs mass by performing the 
derivative of eqs.~(\ref{eq:Pi2L}) and (\ref{eq:Ti2L})  numerically. The numerical derivation in \SPheno is done by using Ridders' method of polynomial extrapolation 
with dynamical step-size \cite{1084580}. We have implemented two different methods to take the derivation of the effective potential which the user can choose as explained later in sec.~\ref{sec:usage}:
\begin{enumerate}
 \item {\bf Purely numerical derivation:} in this approach the entire potential is derived with respect to the VEVs. 
 This is the ansatz of Ref.~\cite{Martin:2002wn}
 \item {\bf Semi-analytical derivation:} in this approach the derivatives of all masses and couplings with respect to the 
 VEVs are calculated separately also in a numerical way. However, all one- and two-loop derivatives of the loop-functions with respect to their arguments have been calculated 
 analytically and implemented in the output \SPheno code. The derivatives of the potential are then calculated combining both results 
 using the chain rule. This can be easily done because every contribution to the potential is a product of couplings ($c_1,c_2$), masses ($m_i$) and a loop function $f_X$ with a coefficient $k$, \qref{eq:Vsimple}.
\end{enumerate}
\begin{subequations}
\label{eq:Vsimple}
\begin{eqnarray}
  V^{(2)}_X &=& k\cdot (c_1 c_2) \cdot f_{X}(m_1^2,m_2^2,m_3^2), \quad \text{for } X=FFS,FFV,SSV\\
  V^{(2)}_X &=& k\cdot (c_1 c_2) \cdot m_{F1} m_{F2} \cdot f_{X}(m_{F1}^2,m_{F2}^2,m_3^2), \quad \text{for } X=\FFSbar,\FFVbar\\
  V^{(2)}_X &=& k\cdot (c_1) \cdot f_{X}(m_1^2,m_2^2), \quad \text{for } X=SS
\end{eqnarray}
\end{subequations}
The second method is numerically slightly more expensive but it is also more stable. In particular, in the 
presence of large hierarchies in the VEVs the purely numerical method could become inaccurate. More details about the 
numerical stability are given in sec.~\ref{sec:usage}.
There is in addition, a third, fully analytical method -- which gives results equivalent to a diagrammatic calculation of the pole mass with external momenta set to zero. We will present the analytic results and implementation in a forthcoming publication \cite{Goodsell:2015ira}. 

\subsection{Calculation of loop corrected mass spectrum}

We have described how the corrections to the effective potential at the two-loop level are calculated
and how to obtain the self-energies of the Higgs from it. We will now show how this fits into the full picture by explaining the different steps
performed in the numerical evaluation by \SPheno to obtain the loop corrected Higgs masses:
\begin{enumerate}
 \item The starting point for all loop calculations is the set of running parameters at the renormalization scale $Q$. This scale can be either 
 be a fixed value or a variable which depends on other parameters of the model. For instance, in SUSY models it is 
 common to choose $Q$ to be the geometric mean of the stop masses.
 \item The running parameters are used to solve the minimisation conditions of the vacuum (the tadpole equations $T_i$) at tree-level
 \begin{equation}
  T_i = \frac{\partial V^{(T)}}{\partial v_i} \equiv 0.
 \end{equation}
These equations are solved for a set of parameters, one per equation. This set is determined by the user; typically these are mass-squared parameters, which can be solved for linearly, but \SARAH also allows non-linear tadpole equations.
\item The running parameters as well as the solutions of the tadpole equations are used to calculate the tree-level mass spectrum. 
The tree-level Higgs masses $m^{h,(T)}_i$ are the eigenvalues of the tree-level mass matrix $M^{h,(T)}$ defined by 
\begin{equation}
 M^{h,(T)} = \frac{\partial^2 V^{(T)}}{\partial \phi_i \partial \phi_j}
\end{equation}
\item Similarly, the tree-level masses of all other particles present in the model are calculated.
\item Using the tree-level masses the one-loop corrections $\delta M_Z$ to the $Z$ boson are calculated
\item The electroweak VEV $v$ is expressed by the measured pole mass of the $Z$, $M_Z^{pole}$, the one-loop corrections and a function of the involved gauge couplings $g_i$. 
\begin{equation}
  v = \sqrt{\frac{M_Z^{2,\text{pole}} + \delta M^2_Z}{f(\{g_i\})}}
\label{eq:electroweakv}\end{equation}
In the case of the MSSM $f(\{g_i\})  = f(g_1, g_2) = \frac{1}{4} (g_1^2 + g_2^2)$ holds. 
Together with the value of the running $\tan\beta$, the values for the VEVs of the up- and down Higgs can be calculated. 
\item The tree-level masses are calculated again with the new values for the VEVs.
\item The one- ($\delta t_i^{(1)}$) and two-loop ($\delta t_i^{(2)}$) corrections to the tadpole equations are calculated. These are used to solve the loop-corrected minimisation conditions 
\begin{equation}
T_i + \delta t_i^{(1)} + \delta  t_i^{(2)} \equiv 0.
\end{equation}
\item The one-loop self-energies for all particles including the external momentum $p$ are calculated. For the Higgs, we call them in the following $\Pi^{h,(1L)}(p^2)$. 
\item For the CP-even Higgs states, the two-loop self-energies (with zero external momentum) $\Pi^{h,(2L)}(0)$ are calculated  as explained in the previous section. 
\item The physical Higgs masses are then calculated by taking the real part of the poles
of the corresponding propagator matrices
\begin{equation}
\mathrm{Det}\left[ p^2_i \mathbf{1} - M^{h,(2L)}(p^2) \right] = 0,
\label{eq:propagator}
\end{equation}
where
\begin{equation}
 M^{2,(2L)}(p^2) = \tilde{M}^{h,(T)} -  \Pi^{h,(1L)}(p^2) - \Pi^{h,(2L)}(0).
\end{equation}
Here, \(\tilde{M}^{h,(T)}\) is the tree-level mass matrix where the parameters solving the loop-corrected tadpole equations are used. 
Eq.~(\ref{eq:propagator}) is solved
for each eigenvalue $p^2=m^2_i$ in an iterative way. 
\end{enumerate}

\subsection{Treating the Goldstones}
\label{sec:goldstones}
It is well known that the derivatives of the effective potential in the Landau gauge may suffer from divergences due to massless Goldstone bosons \cite{Martin:2013gka}: the derivatives of some loop functions have infra-red singularities. In the gaugeless limit of the MSSM, this problem is circumvented because the masses of the pseudoscalars become independent of the Higgs vevs, and so the derivatives of the effective potential do not contain any singular functions. However, once we go beyond the MSSM (even to the NMSSM) this problem reappears. The singularities in the first derivative of the potential may be tamed by resummation methods \cite{Elias-Miro:2014pca,Martin:2014bca}, but the second derivatives may remain singular and it has been suggested that the problem could be resolved by passing to the pole mass calculation instead \cite{Martin:2014bca}. However, working in the gaugeless limit this problem is usually not present: as stated above, the tree-level masses entering the calculation are calculated at the minimum of (full) tree-level potential. Thus, using the Lagrangian 
parameters at this minimum in the mass matrices expressed in the gaugeless limit, the Goldstone masses are usually non-zero and 
no divergences in the numerical evaluation show up. This is one strong motivation for working in the gaugeless limit: it reduces the dependence of the tree-level Goldstone ``masses'' that enter into the calculation on the renormalisation scale; the problem was noted to be particularly severe  for a full calculation of the two-loop Higgs mass via the effective potential technique in the MSSM in \cite{Martin:2002wn}. However, it would be interesting to explore alternative solutions to this problem.

\subsection{Limitations}
As we have stated above, the presented procedure can reproduce the Higgs mass for a wide range of SUSY models with the precision most spectrum generators provide at the moment for the MSSM. That means that we can include all two-loop corrections including the strong coupling and any superpotential or soft-parameter, but neglect those coming from electroweak couplings. However, even within this approximation there is still one remaining correction missing: the two-loop corrections to the electroweak vev $v$. As can be seen from eq.~(\ref{eq:electroweakv}), $v$ will receive corrections of $\frac{\delta M_Z^{2,2L} }{ 2M_Z^{2,\text{pole}} f(\{g_i\})}  $ which in simple extensions of the MSSM will be $\frac{2 \delta M_Z^{2,2L} }{ M_Z^{2,\text{pole}} (g_1^2 + g_2^2)}  $. At any given loop order there will be contributions to  $\delta M_Z^2$ proportional to $g_1^2, g_2^2$ (i.e. loops which do not contain any further electroweak couplings) and so there will be nonvanishing contributions to $v$ even in the limit that the electroweak gauge couplings are set to zero. These corrections then feed into the tree-level Higgs mass; in the MSSM this is not an issue because, in the gaugeless limit, they are multiplied by zero. However, in general extensions (such as the NMSSM) the tree-level Higgs mass matrix will contain non-vanishing elements proportional to $v$, and therefore there will be a corresponding two-loop correction to the Higgs mass. Since the general expression for these two loop corrections is not available in the literature, we leave the calculation and implementation of these to future work. 

A further limitation is that only corrections to CP even states are calculated. Thus, two-loop corrections for pseudo-scalars or charged Higgs 
bosons are not calculated by this setup at the moment.

\section{Validation}
\label{sec:validation}
\begin{figure}[hbt]
\includegraphics[width=0.49\linewidth]{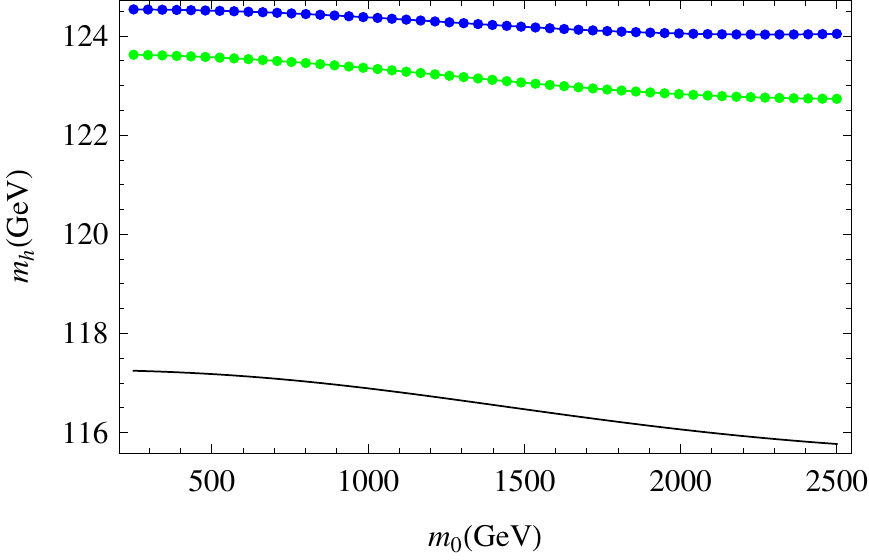} \hfill
\includegraphics[width=0.49\linewidth]{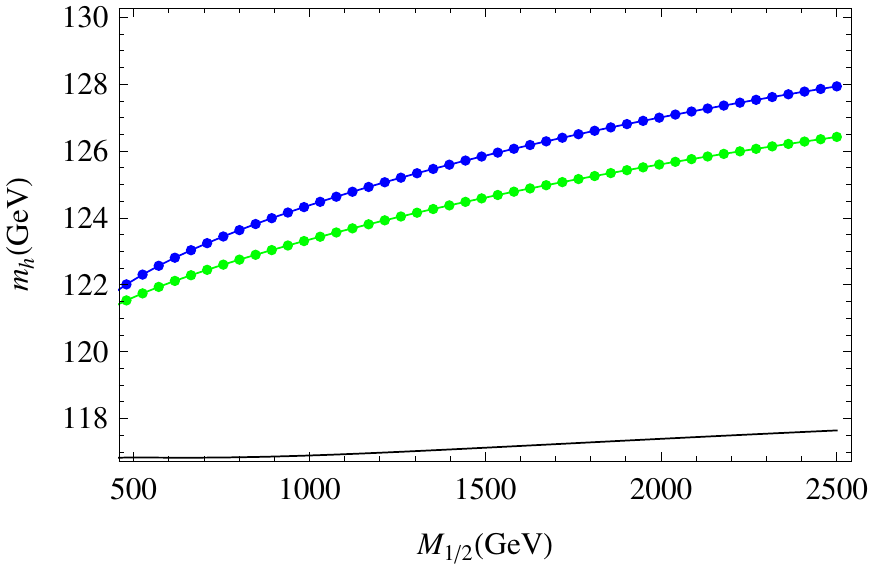} \\
\includegraphics[width=0.49\linewidth]{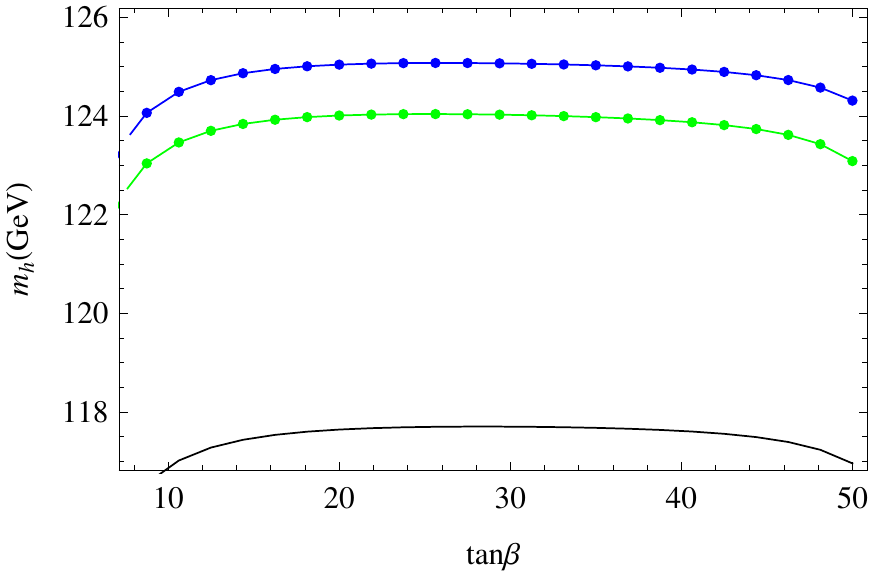} \hfill
\includegraphics[width=0.49\linewidth]{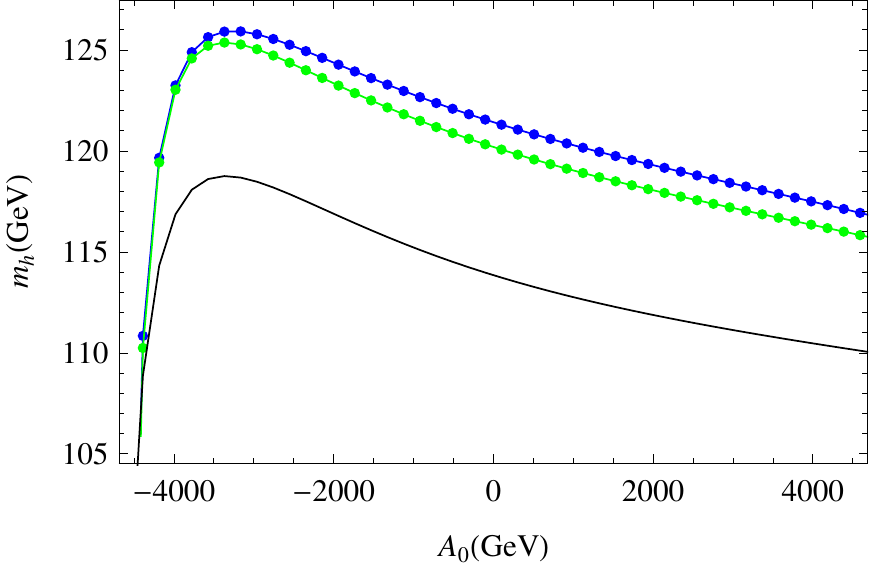}  
\caption{The Higgs mass at one- (black) and two-loop (blue, green) in the CMSSM for a variation of $m_0$ (top left), $M_{1/2}$ (top right),
$\tan\beta$ (bottom left) and $A_0$ (bottom right). The unvaried parameters are fixed to $m_0 = M_{1/2} = 1$~TeV, $\tan\beta=10$, $\mu>0$, $A_0 =-2$~TeV. Blue shows the Higgs mass including $\alpha_s(\alpha_t+\alpha_t)$ corrections while green includes all dominant two-loop corrections. The full lines are the results using the routines of Refs.~\cite{Brignole:2001jy,Degrassi:2001yf,Brignole:2002bz,Dedes:2002dy,Dedes:2003km}, while for the small circles the routines automatically generated by \SARAH are used.}
\label{fig:CMSSM}
\end{figure}

We have cross-checked the Higgs mass using our new routines against the well-established routines of Refs.~\cite{Brignole:2001jy,Degrassi:2001yf,Brignole:2002bz,Dedes:2002dy,Dedes:2003km}. For this purpose, we made a few modifications to ensure that both routines run with equivalent conditions:
\begin{enumerate}
 \item By default the new routine includes any correction from superpotential parameters and soft-terms while the routines of Refs.~\cite{Brignole:2001jy,Degrassi:2001yf,Brignole:2002bz,Dedes:2002dy,Dedes:2003km} are restricted to third generation couplings in the context of the MSSM. Therefore, we set all couplings of the first and second generations of (s)quarks to zero. 
 \item As discussed above, we have implemented a flag to perform calculations in the gaugeless limit in which D-term contributions to the masses are neglected. We made use of this option. However, it was also necessary to ensure that both routines use the same values for the scalar and pseudo-scalar masses: even if $D$-terms in the mass matrices are neglected, the routines of Ref.~\cite{Brignole:2001jy,Degrassi:2001yf,Brignole:2002bz,Dedes:2002dy,Dedes:2003km} are usually called with values for $\mu$ and $M_A$ which correspond to the minimum of the potential {\it including} $D$-terms. Diagonalising the corresponding scalar and pseudo-scalar mass matrix would not give $M_G = M_h = 0$. In our comparison, we therefore used the gaugeless limit and re-solved the tadpole equations in this limit. The obtained values for $\mu$ and $M_A$ were then used in both calculations to ensure that all masses running in the loops are identical.  
\end{enumerate}
The resulting Higgs mass for a variation of $m_0$, $M_{1/2}$, $\tan\beta$ and $A_0$ in the context of the CMSSM is shown in Fig.~\ref{fig:CMSSM} . 
One can see the very good agreement between the automatically generated routines by \SARAH and the ones of Refs.~\cite{Brignole:2001jy,Degrassi:2001yf,Brignole:2002bz,Dedes:2002dy,Dedes:2003km}. 
There are tiny numerical differences stemming from the numerical derivation but those are negligible and have no visible impact on the Higgs mass
as can bee seen from Tab.~\ref{tab:values}. One can also see from these numbers that there is hardly any difference in including $D$-terms to 
the tree-level masses used in the calculation, as expected.

\begin{figure}[hbt]
\includegraphics[width=0.49\linewidth]{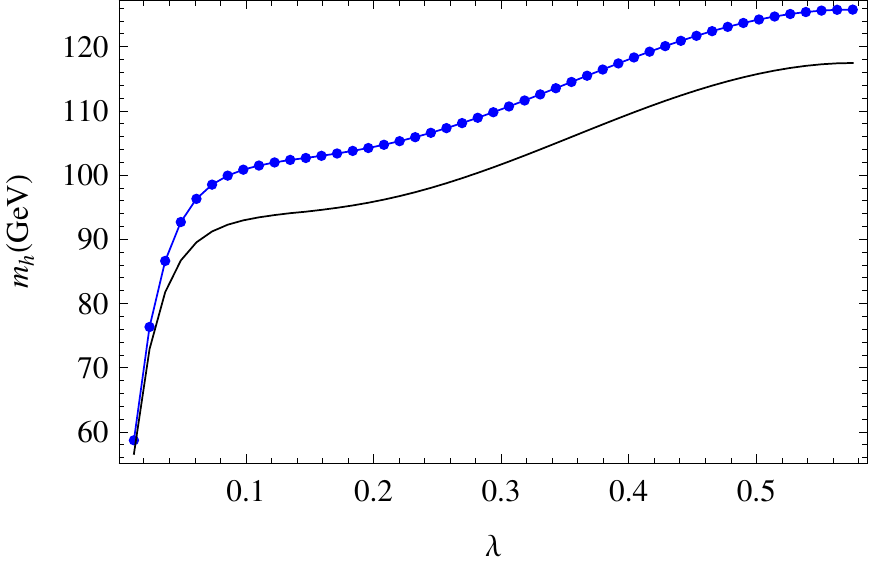} \hfill
 \includegraphics[width=0.49\linewidth]{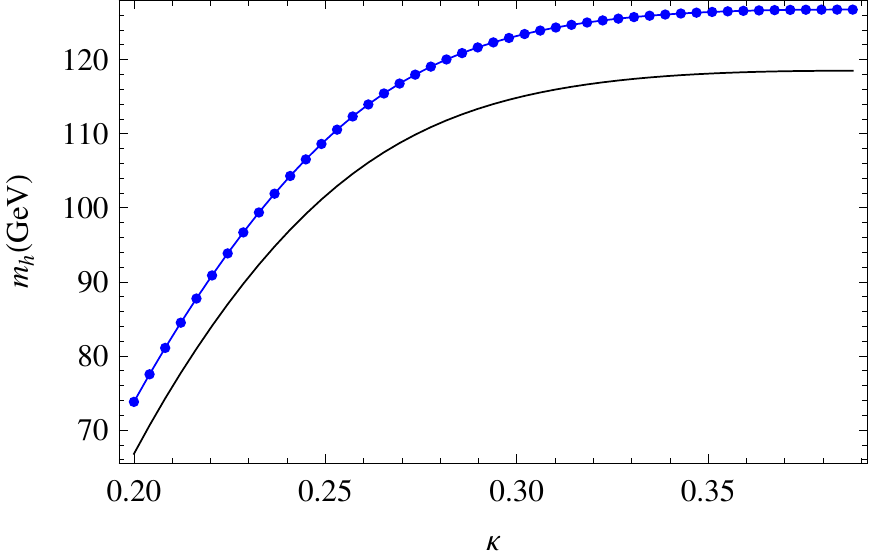} %\\
\caption{The Higgs mass at one- (black) and two-loop with $\alpha_S(\alpha_b+\alpha_t)$ corrections (blue) in a constrained variant of the NMSSM for a variation of $\lambda$ (left) and $\kappa$ (right).  
The unvaried parameters are fixed to $m_0 = M_{1/2} = 1.4$~TeV, $\tan\beta=2.9$, $\mu>0$, $A_0 =-1.35$~TeV, $\lambda=0.56$, $\kappa=0.33$, $A_\lambda=-390$~GeV, $A_\kappa=-280$~GeV, $v_S=500$~GeV.  The full lines are the results using the routines of Ref.~\cite{Degrassi:2009yq}, while for the small circles the automatically generated routines by \SARAH are used.}
\label{fig:NMSSM}
\end{figure}

\begin{table}[hbt]
\begin{tabular}{|c|ccccc|}
\hline  
 & \parbox{2.5cm}{\centering Purely-numerical method}  & \parbox{2.5cm}{\centering Purely-numerical method (gaugeless)} 
 & \parbox{2.5cm}{\centering Semi-analytical method}   & \parbox{2.5cm}{\centering Semi-analytical method (gaugeless)} & Reference \\
\hline 
$\Pi^{(2)}_{11}$ [GeV${}^2$]      &    3475.21 &     3462.95 &    3475.18 &      3462.87   & 3460.45  \\
$\Pi^{(2)}_{12}$ [GeV${}^2$]      &   -299.21  &    -297.92  &   -299.21  &     -297.92    & -297.70  \\
$\Pi^{(2)}_{22}$ [GeV${}^2$]      &    1954.32 &     1954.06 &    1954.32 &      1954.06   & 1954.03  \\
\hline                                                             
$m_{h_1}$ [GeV]       &    124.69  &     124.69  &    124.69  &      124.69    & 124.69   \\
$m_{h_2}$ [GeV]       &    1963.56 &     1963.55 &    1963.56 &      1963.56   & 1963.55  \\
\hline  
\end{tabular}
\caption{Two-loop self energies and loop-corrected masses calculated with the two numerical method to get the 
derivative of the effective potential in the gaugeless limit and with full masses. We used $m_0=M_{1/2}=1$~TeV,
$\mu>0$, $\tan\beta=10$, $A_0 = -2$~TeV. The reference value is the one using the routines of Ref.~\cite{Brignole:2001jy,Degrassi:2001yf,Brignole:2002bz,Dedes:2002dy,Dedes:2003km}.}
\label{tab:values}
\end{table}
As a next step, we compared the two-loop results of the \SARAH routines for other models with existing references: these are the $\alpha_s(\alpha_t + \alpha_b)$ corrections in the NMSSM \cite{Degrassi:2009yq} and in the MSSM extended by Dirac gauginos, for which analytical results and an independent code will be presented in future work. 
Of course, we can use also the new routines to calculate the Higgs masses in these models including the other important corrections, but the presentation and discussion of these results is beyond the scope of this paper and will be given elsewhere \cite{Goodsell:2014pla}. As example for the good agreement between our results and those of Ref.~\cite{Degrassi:2009yq} we show the light Higgs mass in Fig.~\ref{fig:NMSSM} for a variation of $\lambda$ and $\kappa$.

\section{How to use the routines}
\label{sec:usage}
{\tt SARAH 4.4} automatically writes the necessary \Fortran routines in the \SPheno output to calculate the two-loop Higgs corrections. To obtain the \SPheno code for a given model download the most recent \SARAH version from {\tt HepForge} 
\begin{lstlisting}
http://sarah.hepforge.org/ 
\end{lstlisting}
Copy the tar-file into a directory called {\tt \$PATH} in the following and extract it
\begin{lstlisting}
tar -xf SARAH-4.4.0.tar.gz 
\end{lstlisting}
Afterwards, start \Mathematica, load \SARAH, run a  model {\tt \$MODEL} and generated the \SPheno output
\begin{lstlisting}
<< $PATH/SARAH-4.4.0/SARAH.m;
Start["$MODEL"];
MakeSPheno[];
\end{lstlisting}
The last command initializes all necessary calculations and writes all \Fortran files into the output directory of the considered model. These files can be compiled together with \SPheno version 3.3.0 or later. \SPheno is also available at {\tt HepForge}.
\begin{lstlisting}
http://spheno.hepforge.org/ 
\end{lstlisting}
The necessary steps to compile the new files are
\begin{lstlisting}
tar -xf SPheno-3.3.0.tar.gz
cd SPheno-3.3.0
mkdir $MODEL
cp $PATH/SARAH-4.4.0/Output/MODEL/EWSB/SPheno/* MODEL
make Model=$MODEL
\end{lstlisting}
This creates a new binary {\tt bin/SPheno\$MODEL} which reads all input parameters from an external file. 
\SARAH writes a template for this input file which can be used after filling it with numbers as
\begin{lstlisting}
./bin/SPheno$MODEL $MODEL/LesHouches.in.$MODEL 
\end{lstlisting}
The output is written to
\begin{lstlisting}
 SPheno.spc.$MODEL
\end{lstlisting}
and contains all running parameters at the renormalization scale, the loop corrected mass spectrum, two and three-body decays as well as a prediction of precision and flavour observables. More details about the calculation of flavour observables and how to implement new observables are given in the \FlavorKit manual \cite{Porod:2014xia}. The implementation of \SARAH models in \SPheno can also be automatized by 
using the {\tt SUSY Toolbox} \cite{Staub:2011dp}.

There are five flags which can be used in the Les Houches input file to adjust the properties of the two-loop calculation 
\begin{lstlisting}
Block SPhenoInput #
...
  7  ... # Skip two loop masses
  8  ... # Choose two-loop method
  9  ... # Gaugeless limit
 10  ... # Safe mode
...      #
400  ... # Step-size for purely-numerical method
401  ... # Step-size for semi-analytical method
\end{lstlisting}
The following values are possible:
\begin{itemize}
 \item {\tt SPhenoInput[7]}:
 \begin{itemize}
   \item {\tt 0}: Don't skip two-loop masses
   \item {\tt 1}: Skip two-loop masses
  \end{itemize}
 \item {\tt SPhenoInput[8]}:
 \begin{itemize}
   \item {\tt 1}: Two-loop calculation with purely numerical derivation
   \item {\tt 2}: Two-loop calculation with analytical derivation of loop functions (default)
   \item {\tt 9}: Use routines based on Refs.~\cite{Brignole:2001jy,Degrassi:2001yf,Brignole:2002bz,Dedes:2002dy,Dedes:2003km}
  \end{itemize}
 \item {\tt SPhenoInput[9]}:
 \begin{itemize}
   \item {\tt 0}: Turn off gauge-less limit 
   \item {\tt 1}: Use gauge-less limit (default)
 \end{itemize}
  \item {\tt SPhenoInput[10]}:
 \begin{itemize}
   \item {\tt 0}: Turn off the safe-mode (default)
   \item {\tt 1}: Use safe-mode
 \end{itemize}
 \item {\tt SPhenoInput[400]}: a real number (default: 0.5)
 \item {\tt SPhenoInput[401]}: a real number (default: 0.001)
\end{itemize}
Note that the two-loop routines from Refs.~\cite{Brignole:2001jy,Degrassi:2001yf,Brignole:2002bz,Dedes:2002dy,Dedes:2003km} are not included by default 
in the \SPheno output of \SARAH. To include them, in the {\tt SPheno.m} file the flag
\begin{lstlisting}
UseHiggs2LoopMSSM = True;
\end{lstlisting}
has to be set. 

The flags {\tt SPhenoInput[400]} and {\tt SPhenoInput[401]} can be used to check numerical stability of the derivation. If the 
step size is choosen to be too small or large the numerical derivation might suffer from some instabilites. We found that the initial step size for derivation with the fully numerical method usually needs a larger initial step-size for the considered VEV especially for heavy SUSY spectra.
The reason is that the potential is of $O(M_{SUSY}^4)$ and the overall value only changes slightly when the VEVs are varied. 
The second method usually operates acceptably with a smaller initial step size because objects of at most order $O(M_{SUSY}^2,M_Z^2)$ are 
derived numerically. In addition, we make the approximation that in the purely numerical approach mass squareds in the loop which are smaller than $10^{-5}$ times the largest mass squared in the loop is taken to be zero. For the semi-analytical approach we take a limit of $10^{-8}$.
\begin{figure}[hbt]
\centering
\includegraphics[width=0.49\linewidth]{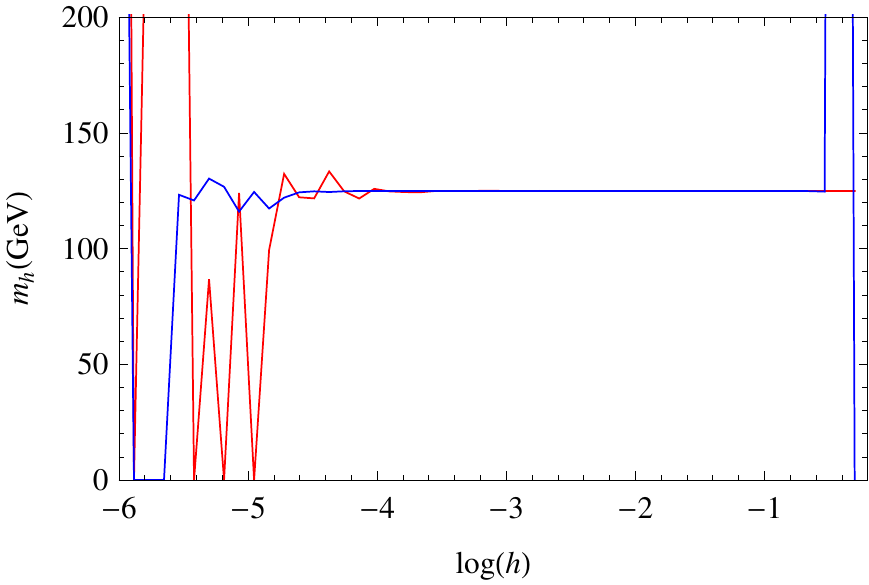}  \hfill
\includegraphics[width=0.49\linewidth]{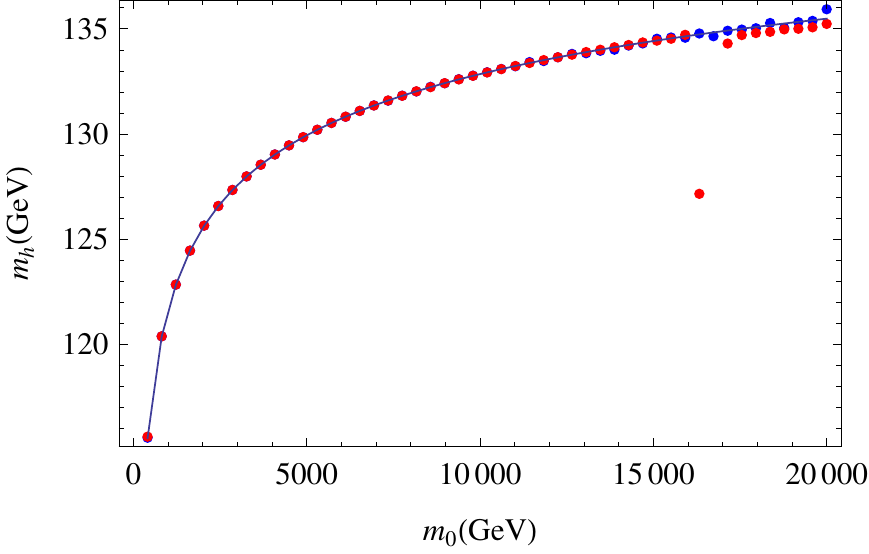}
\caption{Test of the numerical stability: on the left we vary the initial step-size $h$ used in the numerical derivation. The blue line 
corresponds to the semi-analytical approach and red for the purely numerical calculation. On the right we show the Higgs mass as function of a large variation 
of $m_0$ (with $M_{1/2} = -A_0 = m_0$ and $\tan\beta=10,\mu>0$). The solid line is based on the calculation of 
Refs.~\cite{Brignole:2001jy,Degrassi:2001yf,Brignole:2002bz,Dedes:2002dy,Dedes:2003km}, the blue points are based on our semi-analytical 
method and the red ones on the purely numerical one. }
\label{fig:Numerics}
\end{figure}
To give an impression of the numerical stability we show in Fig.~\ref{fig:Numerics} the Higgs mass for a variation of the initial step size 
used in two methods and for a large variation of $m_0$. Here, one sees that the semi-analytical method becomes stable for smaller initial
step sizes as this is the case in the purely numerical calculation. Also we see that the routines are stable even for very large values of the 
SUSY masses. Here, the purely numerical method shows some instabilites for $m_0 > 15$~TeV. This can be improved by changing the initial 
step-size to larger values. The small off-set between the purely-numerical method and the reference in the case of a very heavy SUSY spectrum is explained by our 
approximation to take  ratios of mass squareds smaller than $10^{-5}$ as zero. Because of this even the top quark is treated as massless in some loops.
However, we want to stress that these instabilites 
appear for SUSY masses where this setup should not be used for calculations of Higgs masses. As soon as the SUSY masses are much above the 
EW scale the very heavy particles should be decoupled and an effective theory has to be considered \cite{Bernal:2007uv}. Of course, this statement holds 
for all public versions of SUSY spectrum generators. 

We also provide a "safe mode" for \SPheno via flag 10: in this case \SPheno starts with a large initial step size which is decreased automatically.
It checks for what range of the initial step size the results are numerically stable by comparing the results obtained with different inital step sizes. If no stable range is found an error is returned. In the unlikely case that both methods suffer from numerical instabilities there is also the possibility to increase the numerical precision by passing from double to 
quadruple precision. For this purpose, the Makefile located in {\tt SPheno-3.3.0/src} must be changed. The line
\begin{lstlisting}
PreDef = -DGENERATIONMIXING  -DONLYDOUBLE 
\end{lstlisting}
should be replaced by
\begin{lstlisting}
PreDef = -DGENERATIONMIXING  -DQUADRUPOLE 
\end{lstlisting}
Afterwards the entire \SPheno code must be recompiled via
\begin{lstlisting}
 make cleanall 
 make Model=$MODEL
\end{lstlisting}

This will slow down the numerical evaluation significantly. However, have not found an example for resonable SUSY masses where this was necessary. 

\section{Conclusion and outlook}
\label{sec:conclusion}
We have presented a fully automatised two-loop calculation of Higgs masses in supersymmetric models. The method is based 
on the two-loop effective potential approximation and provides the same numerical accuracy of Higgs masses in beyond MSSM models 
as  commonly used spectrum generators do for the MSSM. That means for any BMSSM model the combination \SARAH/\SPheno
offers now the most precise calculation for Higgs masses available. Of course, there is still much space for 
further improvements: we have not yet included the contributions from electroweak gauge couplings, and also masses for 
CP odd states are not yet calculated; the corrections to $v$ at two loops should be calculated and included; alternative solutions to the Goldstone boson catastrophe should be investigated. However, it is not yet clear if all of these can be accomplished in an appropriate way in the 
effective potential approach. Therefore, future developments will go in the direction of a diagrammatic calculation 
where these effects can be included in a more straightforward way -- a further update of \SARAH together with analytical calculations are in preparation. 
Nevertheless, the presented framework pushes precision studies in non-minimal SUSY models to a new level. It allows
 the prediction of the Higgs mass to be confronted in many SUSY models in the same way as this is done in the MSSM.

\section*{Acknowlegements}
We thank  Stephen Martin for quickly answering our questions concerning Ref.~\cite{Martin:2001vx}. We reserve particular thanks to Pietro Slavich for many informative discussions and pointing out the issue with the two-loop corrections to $v$. 

\begin{appendix}
\section{The Two-loop effective potential in four-component notation} 
\label{app:FourComponentNotation}

The basis for our implementation is the two-loop effective potential for a general renormalizable theory, given by S. Martin \cite{Martin:2001vx}. His convention is the most elegant and simplest when dealing with a general theory, using only Weyl fermions ($\psi_I$), real scalars ($R_K$) and real vectors ($A_\mu^a$). We will refer to it as the R-convention (R for real). 
However, in a specific model it is more useful to organise particles into groups, including explicitly real and complex scalars, Majorana and Dirac fermions. This is the case in the framework of \SARAH/\SPheno : Fermions are described by bispinors $\Psi_i$, which can be Dirac or Majorana, and bosons can be real or complex. We will call this the C-convention (C for complex). In this section we will slightly recast the formulae of the reference to a form suitable for implementation. 

The general structure of the effective potential at two-loop can be decomposed into ten terms,
\begin{eqnarray}
\label{eq:Veff_parts}
  V^{(2)} &=& V_{SSS}^{(2)}+V_{SS}^{(2)}+V_{FFS}^{(2)}+V_{\overline{FF}S}^{(2)}+V_{SSV}^{(2)}+V_{VS}^{(2)}+V_{VVS}^{(2)}+V_{FFV}^{(2)}+V_{\overline{FF}V}^{(2)}+V_{\text{gauge}}^{(2)}.
\end{eqnarray}
Every contribution is described by a simple expression,
\begin{eqnarray}
  V^{(2)} &\sim& (\text{coupling})^2 \times f_{\text{loop}}(m_1^2,m_2^2,m_3^2), \quad \text{sunrise topology}\\
  V^{(2)} &\sim& (\text{coupling}) \times f_{\text{loop}}(m_1^2,m_2^2), \quad \text{snowman topology}
\end{eqnarray}
in which only the prefactors must be figured out carefully. Usually, the loop functions will be abbreviated to $f_{\text{loop}}(m_i^2,m_j^2,m_k^2)=f_{\text{loop}}(i,j,k)$.

In the \SARAH framework, a vertex factor for three scalars $\phi_1,\phi_2,\phi_3$ is understood as
\begin{equation}
  {\text{vertex}} = i\frac{\delta \mathcal L}{\delta \phi_1 \delta \phi_2 \delta \phi_3}=:ic, \label{eq:VertexL}
\end{equation}
which is what is used in the textbook approach to writing down Feynman amplitudes. $\phi_i$ are either Weyl fermions 
or bosons. The vertex factor can be decomposed into a kinematic part (Lorentz indices) and a coefficient. For example, for two Dirac fermions and a vector boson, the vertex factor can be expressed as
\begin{equation}
 {\text{vertex}} = i(c_L P_L+c_R P_R) \gamma^\mu
\end{equation}
Here, $P_L$ and $P_R$ are the polarization operators $P_{L,R} = \frac{1}{2}\left(1\mp \gamma_5\right)$.
The coefficients for particles \verb|p1,p2,p3,(p4)| can be obtained in \SARAH{} by the command 
\begin{lstlisting}
Vertex[{p1,p2,p3}]
Vertex[{p1,p2,p3,p4}]
\end{lstlisting}
E.g. \verb|Vertex[{Fu,bar[Fu],VG}]| returns the vertex of $t,\bar t,g$. 

\SARAH calculates internally with 2-component spinors, but the final set of particles as used in \SPheno is given in terms of Dirac spinors. The Vertex-command accepts both.
If $\xi$ and $\chi$ are 2-component spinors with the same quantum numbers, a Dirac spinor can be constructed (Chiral representation)
\begin{equation}
  \Psi =
  \begin{pmatrix}
    \xi \\ \chi^\dagger
  \end{pmatrix},\ \Psibar = \Psi^\dagger \gamma^0 =(\chi,\xi^\dagger)
\end{equation}
with the translation table
\begin{eqnarray}
 \Psibar_i \gamma^\mu P_L \Psi_j &=& \xi^\dagger_i \bsigmu \xi_j,\qquad \Psibar_i \gamma^\mu P_R \Psi_j = \chi_i \sigma^\mu \chi_j^\dagger \\
 \Psibar_i  P_L \Psi_j &=& \chi_i  \xi_j,\qquad \Psibar_i P_R \Psi_j = \xi_i^\dagger \chi_j^\dagger 
\end{eqnarray}
with $\sigma^\mu=(1,\sigma^i)$ and $\bsigmu=(1,-\sigma^i)$ and the Pauli matrices $\sigma^i$.

For each of the terms in \qref{eq:Veff_parts} we will define a piece of Lagrangian such that $ic$ or $ic_{L/R}$ will match the output of a \SARAH \verb|Vertex| command (cf. eq.~\ref{eq:VertexL}). This Lagrangian is then transformed such that the relations between these couplings and the S.~Martin couplings are obvious. When using \SARAH conventions, we will write lower-case indices $i,j,k$ to denote generation or color indices. When using capital letters $I,J,K$, they refer to the notation where everything is broken down to 2-component spinors $\psi_I=(\xi_1,\chi_1,\xi_2,\chi_2,\dots)$ and real scalars $R_K=(\varphi_1,\sigma_1,\varphi_2,\sigma_2,\dots)$.

\subsection{$FFV$ and $\FFVbar$}
Given a set of fermions $\Psi_i$ and a vector $A^a_\mu$, the Lagrangian term is
\begin{eqnarray}
  \mathcal L_{FFV} &=& \bar \Psi_i \gamma^\mu (c_L P_L + c_R P_R) \Psi_j A^a_\mu \\
&=& c_L \xi^\dagger_i \bar \sigma^\mu \xi_j A^a_\mu + c_R \chi_i \sigma^\mu \chi_j^\dagger A^a_\mu\\
&=& c_L \xi^\dagger_i \bar \sigma^\mu \xi_j A^a_\mu - c_R \chi^\dagger_j \bar\sigma^\mu  \chi_i A^a_\mu \label{eq:LFFV}
\end{eqnarray}
There is {\bf no} implicit sum over $i,j,k$ here. The coefficients are $c_{L/R}=c_{L/R}(i,j,a)$. The Minus sign comes from the rearrangement $\chi \sigma^\mu \xi^\dagger = -\xi^\dagger \bar\sigma^\mu \chi$ (in signatures with mostly plus as well as mostly minus). 
We can arrange all 2-component spinors in a list:  $(\psi_I)=(\xi_1,\chi_1,\xi_2,\chi_2, \dots)$. The interaction in S.~Martin convention is given by
\begin{eqnarray}
  \mathcal L_{FFV} &=& \sum_{I,J,a}g^a_{IJ} \psi_I^\dagger\bsigmu \psi_J A^a_\mu.
\end{eqnarray}
Since this term has to be real, $g^a_{IJ}=(g^a_{JI})^\star$ holds. Vectors only couple left to left and right to right, so a rewriting of \qref{eq:LFFV} is useful:
\begin{eqnarray}
  \mathcal L &=& g^a_{Lij} \xi^\dagger_i \bar \sigma^\mu \xi_j A^a_\mu 
+ g^a_{Rij} \chi^\dagger_i \bar\sigma^\mu  \chi_j A^a_\mu.
\end{eqnarray}
The connection is $g^a_{Lij}=g^a_{I=2i-1,J=2j-1}, g^a_{Rij}=g^a_{I=2i,J=2j}$. This allows to match the coefficients,
\begin{eqnarray}
  g^a_{Lij} &=& c_L(i,j) \\
g^a_{Rij} &=& -c_R(j,i)=-c_R(i,j)^*=-c_R^*.
\end{eqnarray}
The original expression for $\FFVbar$ is given by
\begin{eqnarray}
\label{eq:FFVbar}  V^{(2)}_{\overline{FF}V}&=&\frac 12 g^a_{IJ}g^a_{I^\prime J^\prime} M_{II^\prime} M^\star_{JJ^\prime} F_{\FFVbar}(I,J,a) 
\end{eqnarray}
$M_{IJ}$ denotes a mass insertion, which in case of a Dirac fermion is a $2\times 2$ block matrix, $
\begin{pmatrix}
  0 & m_D \\ m_D & 0 
\end{pmatrix}$, or more formally expressed as $M_{IJ}=m_{Di}(\delta_{I,2i-1}\delta_{J,2i}+\delta_{I,2i}\delta_{J,2i-1})$. 
This allows to partially simplify:
\begin{eqnarray}
  g^a_{IJ}M_{II^\prime}&=&(g^a_{Lij}\delta_{J,2j-1}\delta_{I^\prime,2i}+g^a_{Rij}\delta_{I^\prime,2i-1}\delta_{J,2j})m_i, \\
  g^a_{I\prime J\prime }M_{JJ^\prime}&=&(g^{a}_{Lij}\delta_{I\prime ,2i-1}\delta_{J,2j}+g^{a}_{Rij}\delta_{J,2j-1}\delta_{I^\prime,2i})m_j.
\end{eqnarray}
The whole expression becomes
\begin{eqnarray}
V^{(2)}_{\overline{FF}V}&=&  \frac 12 \sum_{I,J,I^\prime,J^\prime} g^a_{IJ}g^a_{I^\prime J^\prime} M_{II^\prime} M^\star_{JJ^\prime} F_{\overline{FF}V}(m_I^2,m_J^2,m_a^2) \\
&=& \sum_{i,j}g^a_{Lij}g^a_{Rij}m_{Di}m_{Dj}F_{\overline{FF}V}(m_{Di}^2,m_{Dj}^2,m_a^2)
\end{eqnarray}
Observe that for a Dirac fermion, $m_{2i-1}=m_{2i}= m_{Di}$. If we fix two different particles $i$ and $j$, this sum has two terms that involve these particles. 

\begin{eqnarray}
  V^{(2)}_{\overline{FF}V,i\neq j}&=&(g^a_{Lij}g^a_{Rij}+g^a_{Lji}g^a_{Rji})m_{Di}m_{Dj}F_{\overline{FF}V}(i,j,a) \\
&=&2\Re (g^a_{Lij}g^a_{Rij})m_{Di}m_{Dj}F_{\overline{FF}V}(i,j,a)\\
&=&-2\Re (c_L c_R^*) m_{Di}m_{Dj}F_{\overline{FF}V}(i,j,a)
\end{eqnarray}
If we fix the same particle $i=j$, the factor 2 disappears. 
\begin{eqnarray}
  V^{(2)}_{\overline{FF}V,i=j}&=&-\Re (c_L c_R^*) m_{Di}^2F_{\overline{FF}V}(i,i,a)
\end{eqnarray}
Now consider Majorana particles, which are written as
\begin{equation}
  \Psi_{Mi} =
  \begin{pmatrix}
    \xi_i \\ \xi^\dagger_i
  \end{pmatrix}
\end{equation}
For those particles, the mass insertion matrix $M_{IJ}$ has diagonal elements $m_{Mi}$. Lets consider $i$ labelling a Dirac fermion and $j$ a Majorana fermion. Each mass insertion in \qref{eq:FFVbar} can be from each of these fermions. If we take only $M_{II^\str}$ to be the Dirac mass insertions and $M_{JJ^\str}$ the Majorana mass, this gives
\begin{equation}
 g^a_{Lij}g^a_{Rij} m_{Di}m_{Mj} F_{\FFVbar}(i,j,a).
\end{equation}
It can also be the other way round, which leads to
\begin{equation}
 g^a_{Lji}g^a_{Rji} m_{Di}m_{Mj} F_{\FFVbar}(i,j,a).
\end{equation}
In total we have
\begin{eqnarray}
  V^{(2)}_{\FFVbar}&=& 2\Re (g^a_{Lij}g^a_{Rij})m_{Di}m_{Mj} F_{\FFVbar}(i,j,a)\\
&=&-2\Re(c_Lc_R^*) m_{Di}m_{Mj} F_{\FFVbar}(i,j,a).
\end{eqnarray}

When two Majorana fermions interact via gauge coupling, we find $c_R=-c_L^*$, so there is essentially just one coupling $c_L(i,j)$. The indices of the mass insertion $M_{II^\str}$ can again give the mass $m_{Mi}$ or $m_{Mj}$, so \qref{eq:FFVbar} simplifies to 
\begin{eqnarray}
 V^{(2)}_{\FFVbar} &=& \Re((g^a_{Lij})^2) m_{Mi}m_{Mj}F_{\FFVbar}(i,j,a).
\end{eqnarray}
In the case of equal Majorana fermions, $i=j$, we get
\begin{eqnarray}
\label{eq:FFVbarMM}   V^{(2)}_{\FFVbar,i=j} &=& \frac 12 \Re((g^a_{Lii})^2) m_{Mi}^2 F_{\FFVbar}(i,i,a)..
\end{eqnarray}
{\bf Example:} In the case of the gluinos, the gauge interaction term is usually introduced with an $i$, 
\begin{equation}
  \mathcal L_{\text{Glu}} = igf^{abc}\lambda^{a\dagger}\bsigmu A_\mu^b\lambda^c , 
\end{equation}
so here $g^b_{L,ac}=igf^{abc}$, which results in an overall Minus sign in the contribution. $f^{abc}$ are the structure constants of $SU(3)$. Evaluating \qref{eq:FFVbarMM} for this case gives 
\begin{eqnarray}
V^{(2)}_{\spa g}&=&  - \frac 12 g^2 \underbrace{\kl{\sum_{a,b,c=1}^8 (f^{abc})^2}}_{=24} \abs{M_3}^2 F_{\FFVbar}(M_3^2,M_3^2,0)\\
&=& - 12g^2 \abs{M_3^2}F_{\FFVbar}(M_3^2,M_3^2,0)
\end{eqnarray}
This shows the emergence of the color factor of 24. The result matches that of \cite{Martin:2002iu}, Eq.~(3.74). 
Now consider the $FFV$ contributions, given by
\begin{equation}
\label{eq:Veff_FFV}
  V^{(2)}_{FFV} = \frac 12 \sum_{I,J,a}\abs{g^a_{IJ}}^2 F_{FFV}(I,J,a).
\end{equation}
If one of the fermions is Dirac and the other Majorana (MD) (or both Dirac, DD), there are two couplings $g_L,g_R$ involved. For a fixed pair $I\neq J$, there are two equal terms in the sum in \qref{eq:Veff_FFV}. 
\begin{eqnarray}
  V^{(2)}_{FFV,i\neq j} &=& (\abs{g^a_{Lij}}^2+\abs{g^a_{Rij}}^2) F_{FFV}(i,j,a) \quad\text{DD or MD} \\
  V^{(2)}_{FFV,i\neq j} &=& \abs{g^a_{Lij}}^2 F_{FFV}(i,j,a) \quad\text{MM}
\end{eqnarray}
If the fermions are equal, the sum only collects terms $I=J$ and there is a factor of 2 less.
\begin{eqnarray}
  V^{(2)}_{FFV,i=j} &=& \frac 12 (\abs{g^a_{Lii}}^2+\abs{g^a_{Rii}}^2) F_{FFV}(i,i,a) \quad\text{DD} \\
  V^{(2)}_{FFV,i=j} &=& \frac 12 \abs{g^a_{Lii}}^2 F_{FFV}(i,i,a) \quad\text{MM}
\end{eqnarray}
All the different expressions are summarized in Tab.~\ref{tab:FFVFFScollect} together with $FFS/\FFSbar$ discussed next.

\begin{table}
   \centering
  \begin{tabular}{|l|cc|c|cc|} \hline
    & $D_iD_jc$ & $D_iD_jr$ & $D_iM_j$ & $M_i M_jc$ & $M_iM_jr$ \\ \hline
   $V^{(2)}_{FFV},i\neq j$ & $(\abs{c_L}^2+\abs{c_R}^2)$ & $(\abs{c_L}^2+\abs{c_R}^2)$ & $(\abs{c_L}^2+\abs{c_R}^2) $ & $\abs{c}^2 $ & $\abs{c}^2 $\\ 
$V^{(2)}_{FFV},i=j$ & $\frac 12 (\abs{c_L}^2+\abs{c_R}^2)$  &  $\frac 12 (\abs{c_L}^2+\abs{c_R}^2)$ & $-$ &$\frac 12 \abs{c}^2 $ & $\frac 12 \abs{c}^2 $ \\ \hline
   $V^{(2)}_{\FFVbar},i\neq j$ & $-2\Re (c_L c_R^*) $ &$-2\Re (c_L c_R^*) $ &$-2\Re(c_Lc_R^*) $ & $\Re(c^2) $&$\Re(c^2) $ \\ 
$V^{(2)}_{\FFVbar},i=j$ & $-\Re(c_L c_R^*)$ & $-\Re(c_Lc_R^*)$ & $-$ & $-$ & $\frac 12 \Re(c^2)$ \\ \hline \hline
$V^{(2)}_{FFS},i\neq j$ &  $(\abs{c_L}^2+\abs{c_R}^2) $ & $(\abs{c_L}^2+\abs{c_R}^2) $& $(\abs{c_L}^2+\abs{c_R}^2) $ & $\abs{c_L}^2+\abs{c_R}^2 $ & $\abs{c}^2 $\\ 
$V^{(2)}_{FFS},i=j$ & $(\abs{c_L}^2+\abs{c_R}^2) $ & $\frac 12(\abs{c_L}^2+\abs{c_R}^2) $ & $-$ & $\frac 12 (\abs{c_L}^2+\abs{c_R}^2) $& $\frac 12 \abs{c}^2 $\\ \hline
$V^{(2)}_{\FFSbar},i\neq j$ & $2\Re (c_Lc_R^*) $ & $2\Re (c_Lc_R^*) $ &$2\Re (c_Lc_R^*) $ & $2\Re (c_Lc_R^*) $ & $\Re ((c)^2) $\\ 
$V^{(2)}_{\FFSbar},i=j$ & $2\Re (c_Lc_R^*) $ & $\Re (c_Lc_R^*)$ & $-$ & $\Re(c_Lc_R^*)$ & $\frac 12 \Re(c^2)$ \\ \hline \hline
  \end{tabular}
  \caption{Summary of $FFV$, $\FFVbar$, $FFS$, $\FFSbar$ contributions. The contribution is given by $V=k\cdot \text{coup} \cdot f(i,j,k)$ (for $\FFVbar,\FFSbar$ times $m_i m_j$). The table shows the product $k\cdot \text{coup}$ for various cases. $D_iD_jc(r)$ stands for Dirac fermions with complex (real) scalars or vectors, $M_iM_jc(r)$ for Majorana fermions. } 
  \label{tab:FFVFFScollect}
\end{table}

\subsection{$FFS$ and $\FFSbar$}

These contributions are similar in structure to $FFV,\FFVbar$. Consider a set of 4-component fermions $\Psi_i$ and scalars $\phi_k=(\varphi_k+i\sigma_k)/\sqrt{2}$ and constants $c_{L/R}=c_{L/R}(i,j,k)$. Again, for simplicity, consider $i,j,k$ fixed. 
\begin{eqnarray}
  \mathcal L_{FFS} &=&  -\Psibar_i (c_LP_L + c_R P_R) \Psi_j \cdot \phi_k + \hc \\
&=& -( c_L \chi_i\xi_j + c_R \xi_i^\dagger \chi_j^\dagger )\phi_k +\hc \\
&=& -( c_L \chi_i\xi_j \phi_k+ c_R^* \xi_i \chi_j \phi_k^*) +\hc\\
&=& -\kl{ \frac{c_L}{\sqrt 2} \chi_i\xi_jR_k + \frac{ic_L}{\sqrt 2} \chi_i\xi_j\sigma_k +\frac{c_R^*}{\sqrt 2} \chi_j\xi_iR_k + \frac{-ic_R^*}{\sqrt 2} \chi_j\xi_i\sigma_k }  +\hc \label{eq:FFScouplingsDD}
\end{eqnarray}
Note that scalars couple left to right handed parts. In R-convention all scalars are real, labelled as $R_K=(\varphi_1,\sigma_1,\varphi_2,\sigma_2.\dots)$. In this convention, the interaction is given by
\begin{eqnarray}
\label{eq:L_FFS}
  \mathcal L_{FFS} &=& -\frac 12 \sum_{I,J,K}y^{IJK} \psi_I \psi_J R_K + \hc \\
&=& -\frac 12 \sum_{I,K}y^{IIK} (\psi_I) ^2 R_K -\sum_{I< J,\  K}y^{IJK}\psi_I \psi_J R_K + \hc,
\end{eqnarray}
so the coefficient of every term in \qref{eq:FFScouplingsDD} corresponds to a different $y^{IJK}$ with $I<J$. The two-loop contributions to $FFS$ and $\FFSbar$ are
\begin{eqnarray}
\label{eq:Veff_FFS}  V^{(2)}_{FFS} &=& \frac 12 \sum_{I,J,K}\abs{y^{IJK}}^2 F_{FFS}(I,J,K) \\
  V^{(2)}_{\FFSbar} &=& \frac 14\sum_{I,J,K} y^{IJK}y^{I^\str J^\str k} M_{II^\str}^* M_{JJ^\str}^* F_{\FFSbar}(I,J,K) +\hc
\end{eqnarray}
The sum runs freely over $I,J,K$. When evaluating this sum, a symmetry factor of 2 appears in \qref{eq:Veff_FFS} because for each pair $I\neq J$ there is an equal term  with $I,J$ interchanged.
In the $\FFSbar$ case, $I$ can take 4 different indices, each of which give the same expression in the sum.
\begin{eqnarray}
  V^{(2)}_{FFS} &=& \kl{\abs{\frac{c_L}{\sqrt 2}}^2+\abs{\frac{ic_L}{\sqrt 2}}^2+\abs{\frac{c_R^*}{\sqrt 2}}^2+\abs{\frac{-ic_R^*}{\sqrt 2}}^2} F_{FFS}(i,j,k) \\
 \label{eq:FFSresultDD} &=& (\abs{c_L}^2+\abs{c_R}^2)F_{FFS}(i,j,k) \\
  V^{(2)}_{\FFSbar} &=& \kl{ \frac{c_L}{\sqrt 2}\frac{c_R^*}{\sqrt 2}+ \frac{ic_L}{\sqrt 2} \frac{-ic_R^*}{\sqrt 2}} m_{Di}m_{Dj} F_{\FFSbar}(i,j,k) +\text{hc}\\
\label{eq:FFSbarresultsDD} &=&2\Re (c_L c_R^*) m_{Di}m_{Dj} F_{\FFSbar}(i,j,k) 
\end{eqnarray}
If the scalar is real instead of complex, the $\sqrt 2$ will disappear everywhere and $\sigma_k$ can be dropped. This 
leads to the exact same results as eqs.~(\ref{eq:FFSresultDD}) and (\ref{eq:FFSbarresultsDD}).
If there is one Dirac and one Majorana fermion, we can set $\chi_j=\xi_j$ in \qref{eq:FFScouplingsDD}. The result also stays the same, eqs.~(\ref{eq:FFSresultDD}) and (\ref{eq:FFSbarresultsDD}).
Considering two Majorana fermions, i.e. setting $\chi_j=\xi_j$ and $\chi_i=\xi_i$, we get 
\begin{eqnarray}
  \mathcal L_{FFS} &=& -\kl{ \frac{c_L+c_R^*}{\sqrt 2} \xi_i\xi_jR_k + i\frac{c_L-c_R^*}{\sqrt 2} \xi_i\xi_j\sigma_k}  +\hc.
\end{eqnarray}
Evaluating the contribution to the potential gives
\begin{eqnarray}
  V^{(2)}_{FFS} &=& \kl{ \abs{\frac{c_L+c_R^*}{\sqrt 2}}^2+\abs{i\frac{c_L-c_R^*}{\sqrt 2}}^2} F_{FFS}(i,j,k) \\
    &=& \kl{ \abs{c_L}^2+\abs{c_R}^2 } F_{FFS}(i,j,k) \\
V^{(2)}_{\FFSbar} &=& \frac 12 \sum_{I<J}(y^{IJK})^2 m_{Mi}m_{Mj} f_{\FFSbar}(i,j,k) +\hc\\
&=& \frac 12 \kl{\kl{\frac{c_L+c_R^*}{\sqrt 2}}^2+\kl{i\frac{c_L-c_R^*}{\sqrt 2}}^2}m_{Mi}m_{Mj} f_{\FFSbar}(i,j,k) +\hc\\
&=& 2\Re(c_Lc_R^*) m_{Mi}m_{Mj} f_{\FFSbar}(i,j,k)
\end{eqnarray}
If there are two Majoranas and one real scalar, the interaction would be
\begin{eqnarray}
  \mathcal L &=& -\Psibar_i (c_LP_L+c_RP_R) \Psi_j R_k  \\
&=& -c_L \xi_i \xi_j R_k  +c_R \xi_i^\dagger \xi_j^\dagger R_k.
\end{eqnarray}
The complex conjugate is not needed, because the right-handed part already serves that purpose, if $c_R=c_L^*$ is imposed. The contribution to $V^{(2)}$ is
\begin{eqnarray}
  V^{(2)}_{FFS} &=& \abs{c_L}^2 F_{FFS}(i,j,k) \\
  V^{(2)}_{\FFSbar} &=& \frac 12 (c_L)^2 m_{Mi}m_{Mj} F_{\FFSbar}(i,j,k) +\hc \\
&=& \Re((c_L)^2) m_{Mi}m_{Mj} F_{\FFSbar}(i,j,k)
\end{eqnarray}
In the case of equal Dirac fermions and a complex scalar, the interaction Lagrangian will be
\begin{eqnarray}
  \mathcal L_{FFS} &=&  -  \Psibar_i (c_LP_L + c_R P_R) \Psi_i \cdot \phi_k + \hc \\
&=& - \kl{ \frac{c_L+c_R^*}{\sqrt 2} \chi_i\xi_iR_k + \frac{i(c_L-c_R^*)}{\sqrt 2} \chi_i\xi_i\sigma_k}   +\hc, \label{eq:FFScouplingsDDequal}
\end{eqnarray}
which results in 
\begin{eqnarray}
  V^{(2)}_{FFS} &=& (\abs{c_L}^2+\abs{c_R}) F_{FFS}(i,i,k) \\
  V^{(2)}_{\FFSbar} &=& \frac 12 \kl{ \kl{ \frac{c_L+c_R^*}{\sqrt 2}}^2 + \kl{\frac{i(c_L-c_R^*)}{\sqrt 2}}^2} m_{Di}^2 F_{\FFSbar}(i,i,k) +\hc \\
&=& 2\Re (c_Lc_R^*) m_{Di}^2 F_{\FFSbar}(i,i,k).
\end{eqnarray}
In the case of equal Dirac fermions and a real scalar, the interaction Lagrangian is
\begin{eqnarray}
  \mathcal L_{FFS} &=&  -  \Psibar_i (c_LP_L + c_R P_R) \Psi_i \cdot R_k + \hc \\
&=& - \kl{ c_L\chi_i\xi_i + c_R\chi_i^\dagger\xi_i^\dagger}R_k  , \label{eq:FFScouplingsDDequalreal}
\end{eqnarray}
where again $c_R=c_L^*$ is required. This leads to 
\begin{eqnarray}
  V^{(2)}_{FFS} &=& \abs{c_L}^2F_{FFS}(i,i,k) = \frac 12 \kl{\abs{c_L}^2+\abs{c_R}^2} F_{FFS}(i,i,k) \\
  V^{(2)}_{\FFSbar} &=& \frac 12 \kl{ c_L}^2 m_{Di}^2 F_{\FFSbar}(i,i,k) +\hc \\
&=& \Re ((c_L)^2) m_{Di}^2 F_{\FFSbar}(i,i,k)=\Re(c_Lc_R^*)m_{Di}^2F_{\FFSbar}(i,i,k) \\
\end{eqnarray}
Finally, there is the case of equal Majorana fermions and a complex scalar, where we have to start with a factor of $\frac 12$ in $\mathcal L$,
\begin{eqnarray}
  \mathcal L_{FFS} &=& -\frac 12 \Psibar_i (c_LP_L+c_RP_R) \Psi_i \phi_k  +\hc\\
&=& - \frac 12 \kl{c_L \xi_i^2   +c_R (\xi_i^\dagger)^2} \phi_k + \hc \\
&=& -\frac 12 \kl{ \frac{c_L+c_R^*}{\sqrt 2} \xi_i^2 R_k + i\frac{c_L-c_R^*}{\sqrt 2} \xi_i^2\sigma_k}  +\hc.
\end{eqnarray}
This time there is no symmetry factor in the sum over $I,J$, so we end up with
\begin{eqnarray}
  V^{(2)}_{FFS} &=& \frac 12 (\abs{c_L}^2+\abs{c_R}^2)F_{FFS}(i,j,k) \\
  V^{(2)}_{\FFSbar} &=& \frac 14 \kl{ 2c_Lc_R^*} m_{Mi}^2 F_{\FFSbar}(i,i,k) +\hc \\
&=& \Re ((c_Lc_R^*) m_{Mi}^2 F_{\FFSbar}(i,i,k).
\end{eqnarray}
If the scalar is real, we have $c_L=c_R^*$ and
\begin{eqnarray}
  \mathcal L_{FFS} &=& -\frac 12 \Psibar_i (c_LP_L+c_RP_R) \Psi_i R_k  \\
&=& - \frac 12 c_L \xi_i^2 R_k + \hc \\
\Rightarrow V^{(2)}_{FFS} &=& \frac 12 \abs{c_L}^2 F_{FFS}(i,i,k) \\
\Rightarrow V^{(2)}_{\FFSbar} &=& \frac 14 \kl{c_L^2} m_{Mi}^2 F_{\FFSbar}(i,i,k) + \hc \\
&=& \frac 12 \Re(c_L^2)F_{\FFSbar}(i,i,k)
\end{eqnarray}

\subsection{SSS}

In the R-convention this interaction is given by
\begin{eqnarray}
  \mathcal L &=& -\frac 16 \lambda_{ijk} R_i R_j R_k \\
&=& \sum_i (-\frac 16 \lambda_{iii}) R_i^3 + \sum_{i\neq j} (-\frac 12 \lambda_{ijj}) R_i R_j^2 + \sum_{i<j<k} (-\lambda_{ijk}) R_iR_jR_k 
\end{eqnarray}
with three real scalars and $\lambda_{ijk}$ symmetric. The contribution to $V$ can be split up in a similar way,
\begin{eqnarray}
  V_{SSS} &=& \frac 1{12} \sum_{ijk}(\lambda_{ijk})^2 F_{SSS}(i,j,k) \\ 
&=& \sum_{i<j<k} \frac 12 (\lambda_{ijk})^2 F_{SSS}(i,j,k) 
+ \sum_{i\neq j} \frac 14 (\lambda_{ijj})^2 F_{SSS}(i,j,j) 
+ \sum_i \frac 1{12} (\lambda_{iii})^2 F_{SSS}(i,i,i) \label{eq:VSSS}
\end{eqnarray}

Consider complex scalars $\phi_i=(R_i+iI_i)/\sqrt 2$,
\begin{eqnarray}
  \mathcal L&=& c \phi_1 \phi_2 \phi_3 + \text{cc}\\
&=& \frac c{2\sqrt 2} (R_1R_2R_3-(I_1I_2R_3+I_2I_3R_1+I_3I_1R_2)-i(I_1I_2I_3-(R_1R_2I_3+R_2R_3I_1+R_3R_1I_2))) + \text{cc} \\
&=& \frac {c+c^*}{2\sqrt 2} (R_1R_2R_3-(I_1I_2R_3+I_2I_3R_1+I_3I_1R_2)) \\
&+&(-i) \frac {c-c^*}{2\sqrt 2} (I_1I_2I_3-(R_1R_2I_3+R_2R_3I_1+R_3R_1I_2)) \\
&=& \frac {\Re c}{\sqrt 2} (R_1R_2R_3-(I_1I_2R_3+I_2I_3R_1+I_3I_1R_2)) \\
&+& \frac {\Im c}{\sqrt 2} (I_1I_2I_3-(R_1R_2I_3+R_2R_3I_1+R_3R_1I_2)) \\
\end{eqnarray}
Identifying the particles $R_1,R_2,R_3,I_1,I_2,I_3$ with labels $1\dots 6$, we get
\begin{eqnarray}
  \frac {\Re (c)}{\sqrt 2} &=& - \lambda_{123} = \lambda_{453} = \lambda_{561} = \lambda_{642} \\
    \frac {\Im (c)}{\sqrt 2} &=& - \lambda_{456} = \lambda_{126} = \lambda_{234} = \lambda_{315} 
\end{eqnarray}
The effective potential contribution is
\begin{eqnarray}
\nn  V_{SSS} &=& \frac 1{12} (\lambda_{ijk})^2 F_{SSS}(i,j,k) \\ 
\nn &=& \sum_{i<j<k} \frac 12 (\lambda_{ijk})^2 F_{SSS}(i,j,k) 
=   (\Re(c)^2+\Im(c)^2) F_{SSS}(m_1^2,m_2^2,m_3^2)\\
&=&   \abs{c}^2 F_{SSS}(m_1^2,m_2^2,m_3^2) \label{eq:SSSccc}
\end{eqnarray}
Now consider one real scalar, $\phi_3=R_3\in \mathbb R$. 
\begin{eqnarray}
  \mathcal L&=& c \phi_1 \phi_2 \phi_3 + \text{cc} = \frac c{2} (R_1R_2-I_1I_2+i(R_1I_2+R_2I_1))R_3 + \text{cc} \\
&=& \Re(c) (R_1R_2-I_1I_2)R_3 - \Im(c) (R_1I_2+R_2I_1)R_3 
\end{eqnarray}
The remaining five real scalars $(R_1,R_2,R_3,I_1,I_2)$ are labelled $1\dots 5$. 
\begin{eqnarray}
  \Re(c) &=& -\lambda_{123} = \lambda_{345} \\
  \Im(c) &=& \lambda_{134} = \lambda_{234} 
\end{eqnarray}
Plugging this into \qref{eq:VSSS}, we get
\begin{eqnarray}
  V_{SSS} &=& \frac 1{12} (\lambda_{ijk})^2 F_{SSS}(i,j,k) 
= \sum_{i<j<k} \frac 12 (\lambda_{ijk})^2 F_{SSS}(i,j,k) \\
&=&   (\Re(c)^2+\Im(c)^2) F_{SSS}(m_1^2,m_2^2,m_3^2)
=   \abs{c}^2 F_{SSS}(m_1^2,m_2^2,m_3^2)
\end{eqnarray}
which is the same result as \qref{eq:SSSccc}. There is an additional factor of 2 in the coupling, but there are only half the number of independent $\lambda$'s. 
Now, in the case of two real fields $\phi_2,\phi_3$ and one complex field $\phi_1$, 
\begin{eqnarray}
  \mathcal L&=& c \phi_1 \phi_2 \phi_3 + \text{cc}
= \frac c{\sqrt 2} (R_1+iI_1)R_2R_3 + \text{cc} \\
&=& \sqrt 2 \Re(c) R_1R_2R_3 + \sqrt 2 \Im(c) I_1R_2R_3\\
&\Rightarrow& \sqrt 2 \Re(c) = -\lambda_{123} \\
&& \sqrt 2\Im(c) = \lambda_{234},
\end{eqnarray}
there is again a factor of 2 and half the number of real field combinations. The result stays as in \qref{eq:SSSccc}:
\begin{eqnarray}
 V_{SSS} &=& \abs{c}^2 F_{SSS}(m_1^2,m_2^2,m_3^2).
\end{eqnarray}
In the case of three real fields, $c$ is real from the start and $+\text{cc}$ can be omitted. There is only one $\lambda_{123}=-c$, 
\begin{eqnarray}
 V_{SSS} &=& \frac 12 (c)^2 F_{SSS}(m_1^2,m_2^2,m_3^2).
\end{eqnarray}
Now consider two equal complex scalars, $\phi_2=\phi_3$.
\begin{eqnarray}
  \mathcal L&=& \frac c2 \phi_1 \phi_2^2 + \text{cc}\\
&=& \frac {\Re (c)}{2\sqrt 2} (R_1R_2^2-(2I_1I_2R_2+I_2^2R_1)) \\
&+& \frac {\Im (c)}{2\sqrt 2} (I_1I_2^2-(2R_1R_2I_2+R_2^2I_1)) \\
&\Rightarrow& \frac {\Re(c)}{\sqrt 2} = - \lambda_{122} = - \lambda_{155} = \lambda_{245} \\
&& \frac{\Im(c)}{\sqrt 2} = - \lambda_{455} = \lambda_{125} =  \lambda_{224} 
\end{eqnarray}
Plugging this into \qref{eq:VSSS}, we obtain
\begin{eqnarray}
  V_{SSS} &=& \sum_{i\neq j} \frac 14 (\lambda_{ijj})^2 F_{SSS}(i,j,j)+\sum_{i<j<k} \frac 12 (\lambda_{ijk})^2 F_{SSS}(i,j,k) \\
&=& \frac 12  \abs{c}^2 F_{SSS}(m_1^2,m_2^2,m_2^2)
\end{eqnarray}
with a factor of $\frac 12$ compared to \qref{eq:SSSccc}. If $\phi_1=R_1$ is real instead of complex, $\mathcal L$ reads
\begin{eqnarray}
  \mathcal L&=& \frac c2 R_1 \phi_2^2 + \text{cc}\\
&=& \frac {\Re (c)}{2} (R_1R_2^2-R_1I_2^2) - \Im (c) R_1R_2I_2 \\
&\Rightarrow& \Re(c) = - \lambda_{133} = + \lambda_{144} \\
&& \Im(c) = - \lambda_{134} \\
\Rightarrow V_{SSS} &=& \sum_{i\neq j} \frac 14 (\lambda_{ijj})^2 F_{SSS}(i,j,j)+\sum_{i<j<k} \frac 12 (\lambda_{ijk})^2 F_{SSS}(i,j,k) \\
&=& \frac 12 \abs{c}^2 F_{SSS}(m_1^2,m_2^2,m_2^2)
\end{eqnarray}

In the case of two equal real scalars $R_2=R_3$ and one complex scalar $\phi_1$, we get
\begin{eqnarray}
  \mathcal L&=& \frac c2 \phi_1 R_2^2 + \text{cc}\\
&=& \frac {\Re (c)}{\sqrt 2} (R_1R_2^2)
+ \frac {\Im (c)}{\sqrt 2} (I_1I_2^2) \\
&\Rightarrow & \frac {\Re(c)}{\sqrt 2} = -\frac 12 \lambda_{122} ,\quad \frac{\Im(c)}{\sqrt 2} = - \frac 12\lambda_{455} 
\end{eqnarray}
and
\begin{eqnarray}
  V_{SSS} &=& \sum_{i\neq j} \frac 14 (\lambda_{ijj})^2 F_{SSS}(i,j,j)\\
\label{eq:VSSS3R2e} &=& \frac 12  \abs{c}^2 F_{SSS}(m_1^2,m_2^2,m_2^2).
\end{eqnarray}
Turning $\phi_1$ into a real scalar will produce only one term ($\frac c2 R_1R_2^2$) with a real $c=-\lambda_{122}$. This results in 
\begin{eqnarray}
  V_{SSS} &=& \sum_{i\neq j} \frac 14 (\lambda_{ijj})^2 F_{SSS}(i,j,j)\\
\label{eq:VSSS3R3e} &=& \frac 14  \abs{c}^2 F_{SSS}(m_1^2,m_2^2,m_2^2) 
\end{eqnarray}
Consider three equal complex scalars $\phi_1=\phi_2=\phi_3$. 
\begin{eqnarray}
  \mathcal L &=& \frac c6 \phi_1^3 +\text{cc}\\
&=& \frac{\Re(c)}{6\sqrt 2}(R_1^3-3R_1I_1^2) - \frac{\Im(c)}{6\sqrt 2} (3R_1I_1^2-I_1^3) \\
\Rightarrow  \frac{\Re(c)}{\sqrt 2}&= & - \lambda_{111} =  \lambda_{122} \\
 \frac{\Im(c)}{\sqrt 2}&= & - \lambda_{222} =  \lambda_{112} \\
\Rightarrow V_{SSS} &=& \kl{ \frac 18 + \frac 1{24}}\abs{c}^2F_{SSS}(m^2,m^2,m^2)=\frac 16 \abs{c}^2 F_{SSS}(m^2,m^2,m^2)
\end{eqnarray}
At last, if there are three equal real scalars, we get
\begin{eqnarray}
  \mathcal L &=& \frac c6 R_1^3 \\
&\Rightarrow& c = -\lambda_{111} \\
&\Rightarrow&  V_{SSS}= \sum_{i} \frac 1{12} (\lambda_{iii})^2 F_{SSS}(m^2,m^2,m^2).
\end{eqnarray}
All these results are summarized in Tab.~\ref{tab:prefactorsSSS}.

\begin{table}[htpb]
  \centering
  \begin{tabular}{|l|l|l|l|}
    \hline
    fields & all different & two equal & all equal ($\phi_1=\phi_2=\phi_3$) \\ \hline
    $\phi_{1,2,3}\in\mathbb C$ & $1$ & $\nicefrac 12$ & $\nicefrac 16$ \\
    $\phi_{1,2} \in \mathbb C,\phi_3\in \mathbb R$ & $1$ & $\nicefrac 12$ & - \\
    $\phi_{1} \in \mathbb C, \phi_{2,3}\in \mathbb R$ & $1$ & $\nicefrac 12$ & - \\
    $\phi_{1,2,3} \in \mathbb R$ & $\nicefrac 12$ & $\nicefrac 14$ & $\nicefrac 1{12}$ \\  \hline
  \end{tabular}
  \caption{Prefactors for $SSS$ contributions. The contribution is given by $V^{(2)}_{SSS}=k\cdot \abs{c}^2F_{SSS}(m_1^2,m_2^2,m_3^2)$, where $m_i$ is the mass of $\phi_i$. The table shows $k$ for various cases.}
  \label{tab:prefactorsSSS}
\end{table}

\subsection{SS}

The $SS$ contribution is given by
\begin{equation}
  V^{(2)}_{SS} = \frac 18 \sum_{ij} \lambda^{iijj} F_{SS}(m_i^2,m_j^2) \label{eq:VSS}
\end{equation}
In the R-convention this interaction is described by
\begin{equation}
  \mathcal L = - \frac 1{24} \sum_{ijkl}\lambda^{ijkl} R_i R_j R_K R_l
\end{equation}
with a real and completely symmetric $\lambda^{ijkl}$. Picking out only terms where $i=j$ and $k=l$ (both fixed), the sum reads
\begin{equation}
  \mathcal L = - \frac 14 \lambda^{iijj} R_i^2 R_j^2 \quad\text{\bf (no sum, $i\neq j$)}
\end{equation}
If all four scalars are equal, the term is just
\begin{equation}
  \mathcal L = - \frac 1{24} \lambda^{iiii} R_i^4\quad\text{\bf (no sum)}
\end{equation}
Because there are only two scalars in total in the loop, we only have to distinguish the cases of different scalars and equal scalars. In the C-convention with two charged scalars $\phi_1,\phi_2$ we have
\begin{eqnarray}
  \mathcal L &=& c \abs{\phi_1}^2\abs{\phi_2}^2 \\
  \mathcal L &=& \frac c4 \abs{\phi_1}^4 \quad\text{equal scalars}
\end{eqnarray}
where $c$ is the vertex factor in both cases. Introducing $\phi_i=(R_i+i\sigma_i)/\sqrt 2$, this leads to 
\begin{eqnarray}
  \mathcal L &=& \frac c4 \kl{ R_1^2+\sigma_1^2}\kl{R_2^2+\sigma_2^2} \\ 
  \mathcal L &=& \frac c{16} \kl{ R_1^4+\sigma_1^4+2R_1^2\sigma_1^2} \quad\text{equal scalars}
\end{eqnarray}
With this equation, the conventions can be matched. All real scalars $(R_1,R_2,\sigma_1,\sigma_2)$ can be labelled with indices $1,2,3,4$. The coefficients are
\begin{eqnarray}
  -c &=& \lambda^{1122} = \lambda^{1144}=\lambda^{3322}=\lambda^{3344} \quad\text{different scalars} \\ 
  - c &=& \frac 23 \lambda^{1111} = \frac 23 \lambda^{2222} = 2 \lambda^{1122} \quad\text{equal scalars}
\end{eqnarray}
Now simplify the potential contribution \qref{eq:VSS} for complex different scalars. There is a factor of 2 because of symmetry in $i,j$. 
\begin{eqnarray}
  V^{(2)}_{SS} &=& \frac 18 \lambda^{iijj} F_{SS}(i,j) \\
&=& 2\cdot \frac 18 \kl{\lambda^{1122}+\lambda^{1144}+\lambda^{3322}+\lambda^{3344}} F_{SS}(i,j) \\
&=& - c F_{SS}(i,j)
\end{eqnarray}
Now repeat the calculation for two equal complex scalars:
\begin{eqnarray}
  V^{(2)}_{SS} &=& \frac 18 \lambda^{iijj} F_{SS}(i,j) \\
&=& \frac 18 \kl{\lambda^{RRRR}+\lambda^{\sigma\sigma\sigma\sigma}+2\lambda^{RR\sigma\sigma}} F_{SS}(i,i) \\
&=& - \frac 18 \kl{ \frac 32 c + \frac 32 c + c} F_{SS}(i,i) \\
&=& - \frac 12 c F_{SS}(i,i) \label{eq:SSequal}\\
\end{eqnarray}
For one real ($R_2$) and one complex scalar, we get
\begin{eqnarray}
  \mathcal L &=& \frac c2 \abs{\phi_1}^2 R_2^2 \\
  &=& \frac c4 (R_1^2+\sigma_1^2)R_2^2 \\
\Rightarrow -c &=& \lambda^{1122} = \lambda^{2233} \\
\Rightarrow V^{(2)}_{SS} &=& 2 \cdot \frac 18 \kl{\lambda^{1122}+\lambda^{2233}} F_{SS}(i,j)= - \frac c2 F_{SS}(i,j) 
\end{eqnarray}
In the case of two real scalar $R_1,R_2$, we get
\begin{eqnarray}
  \mathcal L &=& \frac c4 R_1^2 R_2^2 \\
\Rightarrow -c &=& \lambda^{1122} \\
\Rightarrow V^{(2)}_{SS} &=& 2 \cdot \frac 18 \kl{\lambda^{1122}} F_{SS}(i,j)= - \frac c4 F_{SS}(i,j) 
\end{eqnarray}
and finally, for two equal real scalars $R_1=R_2$, 
\begin{eqnarray}
  \mathcal L &=& \frac c{24} R_1^4  \\
\Rightarrow -c &=& \lambda^{1111} \\
\Rightarrow V^{(2)}_{SS} &=&  \frac 18 \kl{\lambda^{1111}} F_{SS}(i,i)= - \frac c8 F_{SS}(i,i).
\end{eqnarray}
\begin{table}[htpb]
  \centering
  \begin{tabular}{|l|l|l|}
    \hline
    fields & $\phi_{1,2}$ different & $\phi_1=\phi_2$ equal \\ \hline
    $\phi_{1,2} \in \mathbb C$ & $1$ & $\nicefrac 12$ \\
    $\phi_1 \in \mathbb C, \phi_2 \in \mathbb R$ & $\nicefrac 12$ & - \\
    $\phi_{1,2}\in \mathbb R$ & $\nicefrac 14$ & $\nicefrac 18$ \\ \hline
  \end{tabular}
  \caption{Prefactors for $SS$. The contribution is $V^{(2)}_{SS}=k\cdot (-c)F_{SS}(m_1^2,m_2^2)$. The table shows $k$ for various cases. The masses of $\phi_{1,2}$ are $m_{1,2}$.}
  \label{tab:prefactorsSS}
\end{table}
\subsection{SSV}

In C-convention, the interaction between two complex scalars $\phi_{i},\phi_j$ and a complex vector $W^a_\mu=(A^a_\mu+iB^a_\mu)/\sqrt{2}$ is described by 
\begin{equation}
  \mathcal L_{SSV} = c \phi_i \delmulr \phi_j W^a_\mu + \text{hc}  \label{eq:LSSVC}
\end{equation}
with $c=c(a,i,j)$ ($a,i,j$ fixed) and $f\delmulr g=f\delup\mu g - g \delup\mu f$. The same interaction in the R-convention is given by \cite{Martin:2001vx}, Eq.~(2.12), 
\begin{equation}
  \mathcal L_{SSV} = -\sum_{A,I,J}g^A_{IJ} A^A_\mu R_I \delup\mu R_J
\end{equation}
with $g^A_{IJ}=-g^A_{JI}$, real scalars $R_I$ and real vectors $A^A_\mu$. The potential in R-convention is
\begin{equation}
  \label{eq:VSSV}
  V^{(2)}_{SSV} = \frac 14 \sum_{A,I,J}(g^A_{IJ})^2 F_{SSV}(I,J,A)
\end{equation}
Now break down \qref{eq:LSSVC} to real parts,
\begin{eqnarray}
  \mathcal L_{SSV} &=& c \phi_i \delmulr \phi_j W^a_\mu + \text{hc} \label{eq:LSSVc1}\\
&=& \frac c{2\sqrt 2} (R_i+i\sigma_i)\delmulr (R_j+i\sigma_j)(A^a_\mu+iB^a_\mu)+\hc \\
\nn &=& \frac{\Re({c})}{\sqrt 2} \kl{(R_i\delmulr R_j  - \sigma_i \delmulr \sigma_j) A^a_\mu - (R_i \delmulr \sigma_j+\sigma_i\delmulr R_j) B^a_\mu } \\
 &-& \frac{\Im{(c)}}{\sqrt 2} \kl{(\sigma_i\delmulr R_j + R_i\delmulr \sigma_j) A_\mu^a + (R_i\delmulr R_j-\sigma_i\delmulr\sigma_j)B_\mu^a}.\label{eq:LSSVsplit}
\end{eqnarray}
There are 4 terms for each $\Re(c)$ and $\Im(c)$ which all involve different fields, thus evaluating \qref{eq:VSSV} gives
\begin{eqnarray}
  V^{(2)}_{SSV}&=& 2\cdot\frac 14 \kl{ 4\kl{\frac{\Re(c)}{\sqrt 2}}^2 +4\kl{\frac{\Im(c)}{\sqrt 2}}^2} = \abs{c}^2 F_{SSV}(i,j,a)
\end{eqnarray}
with a symmetry factor of 2 in front because of $g^A_{IJ}=-g^A_{JI}$. If the two scalars are complex conjugates of each other, $\phi_j=\phi_i^*$, \qref{eq:LSSVsplit} reduces to
\begin{eqnarray}
  &\to& \frac{\Re({c})}{\sqrt 2} \kl{ - (R_i \delmulr (-\sigma_i)+\sigma_i\delmulr R_i) B^a_\mu }  - \frac{\Im{(c)}}{\sqrt 2} \kl{(\sigma_i\delmulr R_i + R_i\delmulr (-\sigma_i)) A_\mu^a } \\ 
  &=& \frac{2\Re({c})}{\sqrt 2} \kl{ R_i \delmulr \sigma_i B^a_\mu }  - \frac{2\Im{(c)}}{\sqrt 2} \kl{\sigma_i\delmulr R_i  A_\mu^a } ,
\end{eqnarray}
which gives
\begin{eqnarray}
  V^{(2)}_{SSV}&=& 2\cdot\frac 14 \kl{ 2\kl{\Re(c)}^2 +2\kl{\Im(c)}^2} = \abs{c}^2 F_{SSV}(i,j,a).
\end{eqnarray}
However, if the vector is real and $\phi_i=\phi_j^*$, \qref{eq:LSSVc1} becomes
\begin{eqnarray}
  \mathcal L_{SSV} &=& {ic} (\sigma_i \delmulr R_i) A^a_\mu+\hc,
\end{eqnarray}
where the Hermitean conjugate can be dropped if $c$ is chosen purely imaginary from the start. If that is the case, 
\begin{equation}
  V_{SSV}^{(2)}=\frac{1}{2} \abs{c}^2 F_{SSV}(i,j,a).
\end{equation}
If another field is  considered real, a factor of $\sqrt 2$ disappears in the denominator and we end up with half the terms in $\mathcal L$, which gives again $V_{SSV}^{(2)}=\abs{c}^2F_{SSV}(i,j,a)$. Note that for two real equal scalars, $\mathcal L_{SSV}$ vanishes. All the cases are collected in Tab.~\ref{tab:VSSV}.
\begin{table}[h]
  \centering
  \begin{tabular}[h]{|l|c|} \hline
    fields & $k$ \\ \hline
    $\phi_i,\phi_j,V$ & 1 \\
    $\phi_i=\phi_j^*,V\in \mathbb C$ & 1 \\
    $\phi_i=\phi_j^*,V\in \mathbb R$ & $\nicefrac 12$ \\
    else & 1 \\ \hline
  \end{tabular}
  \caption{This table gives $k$ for the $SSV$ contribution, $V_{SSV}^{(2)}=k\cdot \abs{c}^2 F_{SSV}(i,j,a)$. }
  \label{tab:VSSV}
\end{table}

{\bf Example:} $\spa q_i^{b}, \spa q_i^{c*}, g^a$ with $c=-\frac{g_3}{2}(\lambda^{a})_{cb}$ and $\lambda^a$ the Gell-Mann matrices.
\begin{eqnarray}
\Rightarrow  V^{(2)}_{\spa q_i \spa q^*_i g} &=&  \frac 12 \abs{\frac{g_3}{2}}^2 \underbrace{\kl{\sum_{a=1}^8 \sum_{b,c=1}^3 \abs{\lambda^a_{bc}}^2 }}_{16} F_{SSV}(\spa q_i,\spa q_i,0) \\
&=& 2g_3^2F_{SSV}(\spa q,\spa q,0) \quad \text{c.f. \cite{Martin:2002iu}, (3.48)}.
\end{eqnarray}

\end{appendix}

\bibliographystyle{ArXiv}
\bibliography{lit}

\end{document}